\documentstyle[aaspp4]{article}

\newcommand{\md}{\mbox{$^{\rm o}$}}
\newcommand{\mh}{\mbox{$^{\rm h}$}}

\newcommand{\eg}{e.g.,~}
\newcommand{\ie}{i.e.,~}
\newcommand{\etal}{et al.~}
\newcommand{\kms}{\mbox{$\,$km~s$^{-1}$}}
\newcommand{\hunit}{\mbox{\kms$\,$Mpc$^{-1}$}}
\newcommand{\mpsa}{\mbox{$\,$mag$\,$arcsec$^{-2}$}}
\newcommand{\tot}{\mbox{$_{\rm tot}$}}
\begin{document}

\title{Disk Galaxies in the Outer Local Supercluster: Optical CCD 
       Surface Photometry and Distribution of Galaxy Disk Parameters}
\vspace{0.20in}

\author{Nanyao Y. Lu}
\affil{Infrared Processing and Analysis Center\\
        MS 100-22, Caltech\\
        Pasadena, CA 91125\\
        Email: lu@ipac.caltech.edu}
\vspace{0.20in}

\begin{abstract}
We report new B-band CCD surface photometry on a sample of 76 
disk galaxies brighter than $B_T = 14.5\,$mag in the Uppsala 
General Catalogue of Galaxies, which are confined within a volume
located in the outer part of the Local Supercluster.  With our
earlier published I-band CCD and high S/N-ratio 21cm HI data 
(Lu \etal 1993), this paper completes our optical surface photometry
campaign on this galaxy sample.  As an application of this data
set, the B-band photometry is used here to illustrate two selection
effects which have been somewhat overlooked in the literature, 
but which may be important in deriving the distribution function 
of disk central surface brightness (CSB) of disk galaxies 
from a diameter or/and flux limited sample: a Malmquist-type
bias against disk galaxies with small disk scale lengths (DSL)
at a given CSB; and a disk inclination dependent selection effect
that may, for example, bias toward inclined disks near the threshold
of a diameter limited selection if disks are not completely opaque
in optical.  Taking into consideration these selection effects, 
we present a method of constructing a volume sampling function
and a way to interpret the derived distribution function of CSB
and DSL.  Application of this method to our galaxy sample implies 
that if galaxy disks are {\it optically thin}, CSB and DSL may 
well be correlated in the sense that, up to an inclination-corrected
limiting CSB of about 24.5$\,$\mpsa\ that is adequately probed by
our galaxy sample,  the DSL distribution of galaxies with a lower
CSB may have a longer tail toward large values {\it unless} 
the distribution of disk galaxies as a function of CSB rises rapidly
toward faint values.  
\end{abstract}
\keywords{galaxies: fundamental parameters --- galaxies: photometry 
          --- galaxies: spiral}

\section{Introduction} \label{sec1}

Within the Local Supercluster, multi-color CCD surface photometric
data are now available on a number of flux-limited 
samples of disk galaxies in clusters (\eg Pierce \& Tully 1988; 
Tully \etal 1996), but the same is not true on field disk galaxies.
Lu \etal (1993; hereafter Paper I) selected, in two separate 
volumes in the Local Supercluster (hereafter LSC), all the disk galaxies
brighter than $B_T \approx 14.5\,$mag in the Uppsala General Catalogue 
of Galaxies (hereafter UGC; Nilson 1973).  One of the volumes is
toward but beyond our Local Group as viewed from the Virgo Cluster.
With a median heliocentric velocity of about 2000\kms, the UGC
sample in this ``anti-Virgo Cluster'' volume is located in the outer
part of the Local Supercluster where environmental effects on 
galaxy disks are probably much less important than in and near 
the Virgo Cluster, but still close enough to us to have a fairly
faint absolute magnitude limit of $M_B \sim -17.5\,$mag.  As a 
result, this sample is particularly suitable for probing the statistical
properties of the disk galaxy population.  CCD surface photometry
in the I-band and high S/N-ratio 21cm HI data on this sample are 
already published in Paper~I.  In this paper, we further present 
new CCD surface photometry in the B-band on this sample.

As an application of this data set, the B-band data are used in 
the second half of this paper to probe the distribution function
of galaxy disk parameters.  An exponential disk can be
fully described by two fundamental parameters: a central surface
brightness (hereafter CSB) and an (exponential) disk scale length
(hereafter DSL).  How disk galaxies are distributed in terms of
these two parameters is important as it may carry information 
about the physical condition of the universe at the galaxy formation 
epoch (\eg Dalcanton \etal 1997a).  So far, efforts have been 
mostly in determining this distribution function partially 
integrated over DSL, namely, the CSB distribution.  Freeman (1970)
showed that the CSBs of disks of a sample of local disk galaxies are
distributed in a narrow range of $21.6 \pm 0.3$\mpsa.  This so-called
Freeman law was later interpreted by Disney (1976) and Disney \&
Phillipps (1983) to be a selection effect due to the fact that
a sample selected on the basis of a limiting diameter at a fixed
surface brightness (hereafter SB) may miss giant disk galaxies
(with a large DSL) whose CSBs are too faint, as well as compact
galaxies (with a bright CSB) whose DSLs are too small.  Note 
that the same selection effect could also occur in a flux limited
sample.

The existence of the above selection effect in optical galaxy 
catalogs such as UGC has been supported by a large number of 
studies (\eg Allen \& Shu 1979; Davies~1990; Schombert \etal 1992;
McGaugh \etal 1995;   McGaugh 1996; de Jong 1996; Sprayberry 
\etal 1996;  Dalcanton \etal 1997b; and the references in Bothun
\etal 1997).  But there is still considerable controversy about
the exact shape of the CSB distribution for relatively bright 
disk galaxies.  For example, using a 
diameter-limited galaxy sample and a volume sampling function based
on both CSB and DSL, van der Kruit (1987) concluded that after 
excluding dwarf galaxies,  there are not many large, low-CSB disk
galaxies; on the other hand, using larger galaxy samples and a 
volume sampling function based on CSB alone,  other groups have 
derived CSB distributions that are nearly flat at values fainter
than the canonical Freeman's value of B$\,$21.6\mpsa\ (\eg Davies 
1990; McGaugh \etal 1995; McGaugh 1996).  For a further discussion
on this controversial subject, see Briggs (1997).

There are, however, a number of issues which have not been formally
addressed in the past and which may be important to correctly 
interpreting any CSB distribution function derived from a diameter
or flux limited sample.   The first one concerns 
the so-called Malmquist bias (Malmquist 1920).  At a given CSB,
the DSL distribution is fairly wide (\eg McGaugh \etal 1995,
de~Jong 1996).  A diameter (or flux) limited selection leads
to a Malmquist-type bias in the sense that one tends to select
only intrinsically large disks at a given CSB.  Near the limiting
SB of a sample selection,  a slight dimming in CSB has to be 
compensated by a large increase in DSL in order for a galaxy to
be selected.  Therefore, a bias could occur between high- and
low-CSB disk galaxies.

The second issue is how to take into account the effect of disk 
inclination on the detectability of a disk galaxy.  Depending on 
how transparent a galaxy disk is in optical, this effect could
be particularly important to galaxies with disk parameters near
the selection threshold.  As we show in this paper, depending 
on whether disks are opaque or transparent, the same galaxy 
sample could lead to a quite different conclusion on the derived
distribution function of galaxy disk parameters.

The third issue regards whether the variables, CSB and DSL, are 
separable in the bi-variate galaxy distribution function. In other
words, are CSB and DSL statistically independent of each other?
A positive answer to this question would allow one to derive a 
CSB distribution function from a diameter or flux limited sample
without requiring complete redshift data.  Some authors have 
argued for and used such a statistical independence between CSB
and DSL (\eg McGaugh 1996), but this has never been rigorously
tested.  A possible correlation between the two parameters is 
hinted by theoretical considerations (\eg Dalcanton 1997a) and by 
the observational fact that low-CSB spiral galaxies tend to have
a large DSL (\eg Kent 1985; Bothun \etal 1990; Dalcanton \etal 
1997b).  When examining the apparent DSL distribution of galaxies
from a diameter or flux limited sample, one has to be aware of 
the Malmquist bias:  at fixed CSB and distance, the bias 
prevents one from sampling galaxies with a DSL shorter than some 
threshold.  Without a prior knowledge of the intrinsic DSL 
distribution function, the only secure 
measurement one can make is on the part of the DSL distribution
above this threshold.  Therefore, for a set of well defined 
selection criteria, and as we show in this paper, one can still 
answer the question as to whether the DSL distribution at large 
values depends on CSB.

The remainder of this paper is organized as follows: In Sect.~2, 
we describe the galaxy sample, our B-band CCD surface photometry
and present the photometric results.  In Sect.~3, we illustrate 
the two selection effects introduced above; and by taking into 
consideration these selection effects, present a method of 
constructing a volume sampling function and a way to interpret 
the derived distribution function of disk parameters.  The method
is then applied to our galaxy sample.  In Sect.~4, we discuss 
some implications from our analyses.  We end this paper with a 
brief summary in Sect.~5.  
Throughout this paper, we assume a Hubble constant of 75\hunit, 
and use the following notations for an exponential disk: $\mu_0$
($\mu^c_0$) for its observed (face-on) CSB in units of \mpsa and
$r_s$ ($h_s$) for its angular (linear) DSL in units of arcsec (kpc).
Thus, an exponential disk can be expressed as
$$\mu(r) = \mu_0 + 1.086(r/r_s). \eqno(1)$$

\section{B-Band CCD Surface Photometry} \label{sec2}

\subsection{Observations}\label{sec2.1}

Our galaxy sample, originally selected in Paper I and used in Lu
\etal (1994) to study the local velocity field,  contains all the 76 UGC disk 
galaxies with $B_T < 14.5$ mag within a volume bounded by $22\mh
< \alpha < 2\mh$, $0\md < \delta < 20\md$ and a heliocentric 
velocity of $0 < \upsilon_h < 3000$\kms.   Our B-band CCD observations
were carried out with the Hale $200''$ telescope equipped with 
the four shooter (Gunn \etal 1987) at Palomar Observatory from 
August 25 to 27, 1990 (UT) under a photometric condition.  We used
the 4-shooter's standard violet filter (4300\AA/700\AA) to mimic
Johnson's B system (Johnson \& Morgan 1953).  The resulting CCD field
is a $4'.4$ square with a pixel size of $0.''336$.  In addition to
this UGC sample, we also observed a number of optically fainter 
galaxies as described in Paper~I (also see Hoffman \etal 1996).  
The integration time per galaxy ranges from 4 to 8 minutes.  The data
reduction procedure is similar to that in Paper I.  The final images
ready for surface photometric analysis show a quite flat background.
For example, the mode and mean of sky pixels usually agree with each
other within 0.3\%.   The instrumental magnitudes were converted 
into Johnson B system using the observed standard stars taken from
Landolt (1983).

\subsection{Surface Photometry}\label{sec2.2}

Surface photometric analysis was done by fitting elliptical 
contours to each sky-subtracted galaxy image following the prescription
given in Paper I.  The fitting was performed on the B-band images,
independent of the existing I-band results in Paper I.  In most
cases, a good fit could be obtained down to a SB of B$\,26$\mpsa.
For each galaxy, the resulting SB profile as a function
of the semi-major axis, $r$, was displayed and its outer part between
two radii $r_1$ and $r_2$, dominated by the disk component as judged
by eyes, was fit into eq.~(1).  The mean
ellipticity of the disk component, $e$, was evaluated between 
the radii $r_1$ and $r_2$.   The isophotal magnitude, $B_{26}$, and
diameter, $D_{26}$, were measured at B$\,$26\mpsa\ isophote determined by 
the fitted exponential disk profile.   For a number of galaxies 
whose B$\,26$\mpsa\ isophotes are partially outside the CCD field,
the exponential disk fit was used in each case to evaluate the contribution 
to $B_{26}$ from those isophotes partially outside the CCD field.
Finally, the total magnitude, $B\tot$, was evaluated by 
extrapolating the isophotal magnitude at $r_2$ to infinity in radius
using the exponential disk fit.

We list in Table~1 all the 76 UGC sample galaxies as well as those optically
fainter ones that we observed.  Of the 76 UGC sample galaxies, seven
do not have photometric data for various reasons as given at the end of 
Table~1. The table columns are as follows: Col.~(1) is the galaxy
name as in UGC, but for those fainter galaxies we give their names
as in Paper I.   Col.~(2) gives the NGC or IC number if applicable.
The adopted distance in Mpc is given in col.~(3), derived from 
the velocity of the galaxy with respect to the centroid of the LG 
[\ie $\upsilon_h + 300\sin(l)\cos(b)$].  Cols.~(4) and (5) are $r_1$ 
and $r_2$ in arcsec, respectively; namely, the inner and outer radii
for the exponential disk fit.  Col.~(6) is the mean disk ellipticity,
$e$, which has been used to derive the disk inclination angle in 
degrees in col.~(7) via
$$\cos^2(i) =\cases {{(1-e)^2 - 0.2^2 \over 1 - 0.2^2}, &if $e \le 0.8\md$; \cr
		      0, & otherwise. \cr} \eqno(2)$$   
Col.~(8) is the mean position angle of the disk on the sky (N to E)
measured between the radii $r_1$ and $r_2$.   Cols.~(9) and (10) are 
respectively $B_{26}$ in mag and $D_{26}$ in arcmin.   Col.~(11) is
the angular DSL in arcsec and col.~(12) the linear DSL in kpc.
Col.~(13) is $\mu_0$ in \mpsa\ determined from
the exponential disk fit.  Col.~(14) is the total magnitude $B\tot$.
Col.~(15) is the B-band absolute magnitude derived from $B_{tot}$ 
and the distance in col.~(3).  Finally, Col.~(16) gives $(B-I)$ color
derived from $B\tot$ in this paper and $I\tot$ in Paper~I.  
No correction for Galactic or internal reddening has been applied 
to the parameters in Table~1.

Fig.~1 displays the observed B-band surface brightness as a function
of the semi-major axis for each of the galaxies with photometric 
parameters in Table~1.  The open squares represent the measured 
isophotes, while the filled square represents the fitted isophote at
B$\,26$\mpsa.

\subsection{Uncertainties and Systematics}\label{sec2.3}

A number of galaxies were observed multiple times over the entire
observing run.  The multiple images of the same galaxy were reduced
independently from each other and the results are used as a way 
to measure the statistical uncertainties in the derived photometric
parameters.
Such estimated r.m.s.~uncertainties are on the order of $0.01$ mag
for $B_{26}$, $0.05$ mag for $B\tot$, $3\md$ for the disk inclination
angle, $0.3$\mpsa\ for $\mu_0$, and $7\%$ for $r_s$.  
Another way to illustrate our photometric accuracy 
is to compare the B-band result here with the I-band result in 
Paper I on the same galaxy.  As an example, plotted in Fig.~2 as 
functions of the B-band disk ellipticity are the differences in 
the measured disk position angle (P.A.) and ellipticity between 
the two bandpasses.  As expected, the more inclined a galaxy disk 
is, the better agreement between the two bandpasses is in Fig.~2.   
For galaxies inclined more than $45\md$ ($e \sim 0.3$), 
the typical agreement between the two bandpasses is within $\sim 5\md$
in P.A.~and $\sim 10\%$ in disk ellipticity (or $\sim 2.5\md$ in 
terms of disk inclination angle).

Our $B\tot$ magnitudes are however fainter by about 0.06 mag on 
average than the $B_T$ magnitude scale of the Third Reference 
Catalogue of Galaxies (RC3; de Vaucouleurs \etal 1991).  This is
illustrated in Fig.~3.  A Gaussian curve, with a center at $(B\tot-B_T)
= 0.06$ mag and a FWHM of $0.4$ mag, is shown in the figure for
comparison.   No obvious correlation could be identified between 
$(B\tot - B_T)$ and the night on which $B\tot$ was obtained, galaxy 
morphology, optical color or $B_T$.

It is also interesting to see how the fraction of light outside 
the B$\,$26\mpsa\ isophote vary with the disk central surface
brightness. Fig.~4 shows that $(B_{26} - B\tot)$ increases as 
$\mu_0$ increases.  Note that, for a low-SB galaxy of $\mu_0 
\gtrsim 24$\mpsa, more than half of its luminosity lies outside
the B$\,26$\mpsa\ isophote.

Because we did not do a full bulge/disk decomposition, the CSB 
of a galaxy with a prominent bulge could be overestimated 
(Kormendy 1977).   We found however that the eraly-type disk
galaxies do not show on average a brighter CSB than
those late-type galaxies, suggesting that our disk fitting procedure
is probably insignificantly affected by the size of a galactic 
bulge.   On the other hand, there are a number of sample galaxies
with prominent spiral arms forming a ring-like pattern.  These include 
UGC$\,$12343, UGC$\,$12447, UGC$\,$12777, and UGC$\,$00858. For
each of these galaxies, the SB profile outside the spiral arms, 
where we have fit its exponential disk, radially falls off fairly 
quickly to a faint level.  The resulting CSBs of these galaxies
are among the brightest in the sample.   Should we have fit an
exponential disk to the entire galaxy surface,  we would have 
obtained a fainter CSB in each galaxy.  It is not clear which way 
is better.  But not all high-SB sample galaxies are of this type. 
For example,  UGC$\,$12074, UGC$\,$12529 and UGC$\,$00167 are also
among those of the brightest CSBs in the sample, but none of them
show prominent spiral arms in optical.  In fact, with $18.3 < \mu_0
< 19.5$\mpsa\ and a moderate disk inclination, these 3 disk galaxies
may represent a class of relatively rare, ``super high-SB'' disk 
galaxies.  We will study these three galaxies in more details in
a future paper.

\section{On the Disk Parameter Distribution Function} \label{sec3}

\subsection{Formulation of the Sample Selection} \label{sec3.1}

As a conventional simplification, we formulate the UGC sample 
selection by assuming a negligible effect from the bulge of 
a galaxy.  This is probably a reasonable simplification as most 
of our sample galaxies are dominated by their disks.  
We define an intrinsic or face-on CSB as follows:
$$ \mu_0^c = \cases {\mu_0 - 2.5\,K\log (1-e), &if $e \le 0.8\md$; \cr
                     \mu_0 - 2.5\,K\log (1-0.8), & othersie. \cr} \eqno(3)$$
The transition at $e = 0.8$ in eq.~(3) is chosen somewhat arbitrarily.
It corresponds to the onset of $i = 90\md$ when the disk inclination
angle $i$ is given by eq.~(2).  The value of $K$ is limited to $0 
\le K \le 1$, with the lower limit corresponding to a completely 
opaque disk and the upper limit to a fully transparent disk.  In spite
of extensive studies on the opacities of galaxy disks, it is still 
highly controversial as to whether galaxy disks are largely opaque 
or transparent (\eg Tully \& Fouqu\'e 1985; Disney, Davies, \& 
Phillipps 1989; Valentijn 1990, 1994; Burstein, Haynes, \& Faber 1991;
Byun 1993; Giovanelli \etal 1994; Xu \& Buat 1995; Tully \& 
Verheijen 1997).  In this paper we consider only two cases: (a) fully
transparent disks with $K = 1.0$ and (b) fairly opaque disks with 
$K = 0.2$.

We plot in Fig.~5 $\mu_0$ as a function of $r_s$ for the 69 UGC sample
galaxies with B-band photometry. The filled and open squares represent
those with a disk ellipticity below and above $0.5$, respectively. 
It is clear that, at a given $r_s$, disks of larger inclination 
angles are on average associated with brighter values of $\mu_0$
[cf. inequality~(6) below]. 
The solid curve in the figure represents the selection limit associated
with the UGC limiting diameter of $1'$ at $\mu_B \approx 25.3\mpsa$ 
(Cornell \etal 1987) as follows:
$$\mu_0 \le 25.3 - 1.086(30''/r_s). \eqno(4)$$
For an exponential disk, its total magnitude can be written as
$$ B_T = \mu_0 - 5\log r_s - 2.5\log(1-e) - 1.995. \eqno(5)$$
Our magnitude selection criterion of $B_T \le 14.5$ mag transfers to 
$$ \mu_0 \le 5\log(r_s) + 2.5\log (1-e) + 16.49. \eqno(6)$$
Note that as in eq.~(3), we simply set $e = 0.8$ in both criteria~(5)
and (6) for cases of $e > 0.8$.  It is clear that only for a 
fully transparent disk, is criterion~(5) independent of $e$.  We plot
criterion~(6) in Fig.~5 for the cases of $e = 0$ and $e = 0.8$ by the 
dotted and dashed curves, respectively.

To have a rough, but quantitative picture of the overall sample 
selection, we give in Table~2 a few numerical indicators on how
our sample selection acts on {\it face-on} disks of a given CSB: 
Column~(2) is the minimum
angular DSL for a galaxy of a given CSB to be selected.  This is 
given by combining criteria~(4) and (6).   Column~(3) gives 
$\Gamma(-17.5)/\Gamma^s_{max}$,  where $\Gamma^s_{max}$ is the
sample distance cutoff and is taken to be 42.5~Mpc in this paper,
and $\Gamma(-17.5)$ is the maximum distance at which a galaxy of
$M_B = -17.5\,$mag can still be selected.  This quantity
is given by relating the minimum $r_s$ in column~(2) to the following
relation on the absolute magnitude in the case of $M_B = -17.5\,$mag 
and $e = 0$,
$$M_B = \mu_0 - 5\log(h_s) - 2.5\log (1-e) - 38.57. \eqno(7)$$
Column~(3) shows that  a disk galaxy of $M_B = -17.5$~mag
can be seen up to about half of $\Gamma^s_{max}$ in distance for the most 
part of the CSB range explored here.  So roughly speaking, the part
of the galaxy population with $M_B < -17.5\,$mag is adequately
sampled here.  Finally, columns~(4) and (5) are $h_s(-17.5)$ and
$h_s(\Gamma^s_{max})$, respectively, where $h_s(-17.5)$ is the linear
DSL of a galaxy of $M_B = -17.5\,$mag via eq.~(7) and 
$h_s(\Gamma^s_{max})$ is the minimum $h_s$ that a disk galaxy 
has to have in order to be selected out to the sample cutoff 
distance $\Gamma^s_{max}$.

\subsection{Illustration of a Malmquist Bias}\label{sec3.2}

The presence of a Malmquist bias in our sample is illustrated in 
Fig.~6, where we plot $\mu_0^c$ as function of $h_s$ for the sample
galaxies in two cases: (a) fully transparent disks with $K=1.0$,
and (b) fairly opaque disks with $K=0.2$.  The dotted and solid
curves in the figure are generated by using columns~(4) and (5)
of Table~2, respectively.  Clearly, 
the distribution pattern of the data points in each plot suggests
that at a given CSB, only galaxies with large enough $h_s$ have
been selected.   This bias becomes progressively severe when 
the CSB under consideration approaches the limiting SB of 
the sample selection.  Without taking into consideration this
bias,  a distribution of CSB or DSL derived from our UGC sample
would also be biased.   Unfortunately, one usually does not have 
a prior knowledge about the intrinsic shape of the $h_s$ distribution,
especially at a faint CSB level, a correction for this Malmquist 
bias remains model dependent at best.

Although this Malmquist bias makes it impossible to use our sample
to conduct a full bi-variate analysis of the galaxy distribution, 
we can still determine, at a given CSB, the part of the $h_s$ 
distribution that is adequately sampled by our sample.  Roughly 
speaking, this is the region to the right of each dotted curve in
Fig.~7.

\subsection{Derivation of a Volume Sampling Function}\label{sec3.3}

Let us consider a galaxy of an exponential disk with fixed $\mu_0^c$
and $h_s$.  The corresponding observables are $\mu_0$ and $r_s$, 
respectively.  Such a 
galaxy would be selected if its distance $\Gamma$, which relates
$h_s$ to $r_s$, and inclination angle $i$, which relates $\mu_0^c$
to $\mu_0$ via eq.~(3), are such that both criteria~(4) and (6) are 
satisfied.  We sketch this in Fig.~7 for both the cases of $K =1.0$
and $K=0.2$, with the help from the same curves as shown in Fig.~5.
For a fully transparent disk with $K=1.0$, at a given $r_s$, an 
increasing disk inclination would move the galaxy vertically upward,
as illustrated by the thick arrow, from the horizontal line marked as 
``$i = 0\md$'' to the one marked as ``$i\ge 80\md$ (K=1.0).''  This
remains true as long as $r_s$ is greater than that of the vertical
line ``a - b$_1$''  in the figure, which marks the farthest
point in distance at which this galaxy can still be selected in 
our UGC sample.  For a fairly opaque disk with $K=0.2$, at a large
value of $r_s$, an increasing disk inclination still moves the galaxy
vertically upward, but only to the horizontal line marked as ``$i \ge 80\md$ 
(K=0.2).''  The situation changes when the $r_s$ value of the galaxy
becomes smaller than that of point ``b2'' in Fig.~7.  When this happens,
at each $r_s$, the galaxy can move vertically up to the line ``a-b$_2$''
as illustrated by the thin arrow in Fig.~7.   In other words, 
the galaxy would be selected only if its disk inclination is small
enough, and at point ``a'' the galaxy would be included in our UGC
sample only if its disk is viewed face-on.   To summarize, 
the detectability of an optically thin disk is much independent of
its inclination, while a fairly opaque disk could be selected farther
in distance at smaller disk inclination angle.  This statement 
needs a slight modification when the solid curve surpasses both
the dashed and dotted curves in Fig.~7 and becomes the most stringent
selection criterion, \ie at $\mu_0 < 18$\mpsa\ or $\mu_0 \gtrsim 
25$\mpsa.  But as evident in Fig.~5, few sample galaxies are in these
regimes.

We now incorporate this inclination dependent detectability into
a volume sampling function.  For a given galaxy at a given distance
$\Gamma$ (or $r_s$), one can define $V^i_{max}$, the maximum 
``volume'' in the phase space of the disk inclination angle, to be 
$$V^i_{max} = (i_{max} - i_{min})/90\md, \eqno(8)$$
where $i_{min}$ and $i_{max}$, both in degrees, are the minimum 
and maximum possible values for the inclination angle of this galaxy
as described above.   $i_{min}$ can be either zero or greater.  For
$K = 1.0$, $i_{max}$ always equals $90$\md.  For $K < 1$, $i_{max}$ 
could be either $90\md$ or smaller.

Next we let the distance of the galaxy vary (so does $r_s$),  both 
$i_{min}$ and $i_{max}$ are now functions of the distance of 
the galaxy.  Denote $\Gamma^i_{max}$ as the maximum distance a galaxy
can still be detected when it is at the most favorable disk inclination
(\eg point ``a'' in Fig.~7),  we can define a composite maximum 
(space $+$ inclination) phase volume as follows
$$V_{max} = \int_0^{min(\Gamma^i_{max},
  \Gamma^s_{max})}V^i_{max}\Omega\,\Gamma^2\,d\Gamma, \eqno(9)$$
where $\Omega$ is the constant solid angle ($\approx 0.13\,$sr) 
of our sample on the sky, and $\Gamma^s_{max}$ ($= 42.5\,$Mpc) is 
our sample distance cutoff. 

Let $n(\mu_0^c, h_s)$ be the space density distribution function
of disk galaxies in terms of CSB and DSL, in the absence of 
the Malmquist bias, one would have
$$ n(\mu_0^c, h_s)\Delta\mu_0^c\Delta h_s  = \Sigma\,(1/V_{max}), 
\eqno(10)$$
where the sum is over all the sample galaxies with $\mu_0^c$ and 
$h_s$ within the intervals $\Delta\mu_0^c$ and $\Delta h_s$, 
respectively.

\subsection{Results from our UGC Galaxy Sample}\label{sec3.4}

We have calculated $V_{max}$ for each of the 69 UGC sample galaxies
with the B photometry in Table~1.  Although we left out the other 
7 sample galaxies because of their unavailable B photometry, this 
should not have a significant effect on the shape of our derived 
distribution function.  The Malmquist bias and our moderate 
sample size prevent us from using eq.~(10) directly. Instead, we
divide the UGC sample into 4 consecutive bins in the inclination
corrected CSB and derive a DSL distribution within each of these 
bins.   The results are shown in Fig.~8 for the case of $K = 1.0$
and in Fig.~9 for the case of $K = 0.2$.   One galaxy (UGC$\,$01466)
with $h_s \approx 12\,$kpc and a relative density of $\sim 0.04$
for the bin of $21.5 < \mu_0^c < 23.0\,$\mpsa\ is off the figure.
As noted in Table~1, the disk fit of this galaxy was performed 
over a range of radii dominated by the prominent spiral arms of 
the galaxy.  It is likely that its DSL has been overestimated.
A total of 10 sample galaxies with $M_B > -17.5\,$mag have also been
excluded from these figures.   Should they be included here, most of
them would occupy the region in each CSB bin to the left of the 
arrow which roughly indicates the threshold in $h_s$ below which
the Malmquist bias makes the distribution incomplete (see Fig.~6
or Table~2).  The part of the distribution to the right of the arrow
is considered here to represent the true DSL distribution subject
to the statistical error.  It is this part of the distribution we
use to draw the following results.

Let us concentrate on the two faintest, equally wide CSB bins in these 
figures,  namely, (i) $21.5 < \mu_0^c < 23.0\,$\mpsa and (ii) $23.0 <
\mu_0^c < 24.5\,$\mpsa.  These bins cover the flat part of the CSB distribution
shown, for example, in McGaugh (1996).  In the case of transparent
disks with $K=1.0$ (see Fig.~8),  the distribution in the fainter CSB 
bin (ii) has a longer tail toward large $h_s$ values.  For example,  
the integrated density over $h_s > 5\,$kpc is about $0.41$($\pm 
0.17$) in bin (ii).  Note that all the galaxies with $h_s > 5\,$kpc 
in bin (i) would be detectable up to the maximum sample distance 
(cf. Table~2).  Should the CSB distribution be nearly flat with CSB 
and DSL being statistically independent of each other (\eg McGaugh 
1996), one would expect to see about $11$($\pm 4$) [$\approx 0.41 
\times (1/3) \times (\Gamma^s_{max}/10\,{\rm Mpc})^3$] galaxies
with $h_s > 5\,$kpc in bin (i).  But we actually observed only one 
galaxy.  Thus, at a significance level of 2.5$\sigma$, our analysis 
suggests either (a) that CSB and DSL are correlated in the sense 
that the DSL distribution at a fainter CSB level has a longer tail 
toward large values; or (b) that CSB and DSL are still independent 
of each other, but with a CSB distribution function that increases
rapidly toward faint CSB values (\ie at a rate 10 times faster than
a flat CSB distribution).

In the case of fairly opaque disks with $K = 0.2$ as shown in Fig.~9,
we have only marginally useful data in the faintest CSB bin (ii).
Under the assumptions of a nearly flat CSB distribution and a statistical
independence between CSB and DSL,  the same analysis as above implies
that the expected number of galaxies with $h_s > 5\,$kpc in the CSB
bin (i) is $5\,$($\pm 3$).  We actually observed 5 galaxies in that
bin.  However, this comparison is only significant at 1.5$\sigma$.

\section{Discussion} \label{sec4}

The two selection effects that we discussed and formulated in the
previous section, namely, a Malmquist bias and a disk inclination
dependent detectability, should be present in any diameter or/and
flux limited sample.   As we have shown here, both of these could
have a significant effect on how to interpret a distribution 
function derived from a diameter/magnitude limited sample.  We 
note that neither the visibility theory of Disney and Phillipps 
(1983) nor the volume sampling function of McGaugh (1996) has 
fully addressed these selection effects.

For a diameter limited sample, the volume sampling function is 
proportional to $h_s^3(\mu_{limit} - \mu_0)^3$ (McGaugh 1996),
where $\mu_{limit}$ is the limiting SB in the sample selection.
This sensitive dependence on $h_s$ makes the Malmquist bias 
particularly severe near the limiting SB of a sample selection. 
For example, our Fig.~6 shows that $h_s^{min}$,  the threshold
in $h_s$ below which our UGC sample is highly incomplete, 
increases from about 1.5$\,$kpc for the range of $21.5 < \mu_0^c 
< 23\,$\mpsa\ to 3.5$\,$kpc for $23 < \mu_0^c < 24.5\,$\mpsa.
This Malmquist bias, if not corrected for, has the following 
implications:   If the full volume sampling function as defined
above is used (see an example in van der Kruit 1987), one will 
always overestimate the {\it mean} volume sampling function for
the population of the low-CSB galaxies relative to that of 
the high-CSB galaxies, leading to a relative underestimate of
the galaxy number density at low CSB values.  On the other hand,
if a volume sampling function that scales only with $(\mu_{limit}
- \mu_0)^3$ is used (see an example in McGaugh 1996),  one will 
relatively underestimate the volume sampling functions for the
low-CSB galaxies in the sample.  Although this underestimate works
in favor of compensating the Malmquist bias, it is no guarantee 
that the compensation would work out perfectly.

To better understand how disk parameters are distributed at the low
SB end and to effectively constrain disk formation models such as
that proposed by Dalcanton \etal (1997a), we need galaxy samples
selected with a small limiting diameter at a faint SB.
Unfortunately, this implies that the resulting catalog might be 
too large for achieving a completeness in redshift.  One 
alternative way is to use some cluster sample as illustrated
in Tully \& Verheijen (1997), for which complete redshift data
are not needed (except for weeding out field galaxy contaminations).
Note that the selection effects discussed in this paper still
apply to a cluster sample.  A good example of this is shown
in Fig.~4 of Dalcanton \etal (1997a) on a complete sample of 
Virgo Cluster galaxies.  A systematic HI survey (\eg Szomoru \etal
1994) with follow-up optical CCD photometry on the detected galaxies
is another alternative approach that could offer an unbiased picture 
on how galaxy disk parameters distribute above a certain threshold 
in HI mass.

Although the shape of the CSB distribution function is still quite 
uncertain at faint SB levels, it is rather clear from this study 
and those cited in Sect.~1 that this distribution function is fairly 
wide over CSB even for bright disk galaxies, extending to much fainter
values than the narrow range initially proposed by Freeman.  However,
our data (see Figs.~5 and 6) do not show a clear support for a bimodal
CSB distribution.  This is further confirmed by our I-band data from
Paper I in which disk internal reddening is only about 40\% of that
in the B-band.  A bimodal CSB distribution is observed on disk galaxies
in the Ursa Major cluster in both I and K (Tully \& Verheijen
1997).  It is not clear at this point if this difference implies 
that the bimodal CSB distribution is a phenomenon specific to certain
clusters.

\section{Summary} \label{sec5}

In this paper, we present new B-band CCD surface photometry on 
(1) 69 galaxies in a complete sample of 76 disk galaxies brighter
than $B_T = 14.5\,$mag in the Uppsala General Catalogue of Galaxies
and (2) 11 additional galaxies that are optically fainter than 
14.5$\,$mag.  Surface brightness and color radial profiles are 
shown and various photometric parameters are tabulated on each 
of these galaxies.  This data set complements our earlier published
I-band CCD and high S/N-ratio 21cm HI data on the same galaxies
(Lu \etal 1993).

The B-band data are then used to study the distribution of the
fundamental galaxy disk parameters: the central surface brightness
(CSB) and (exponential) disk scale lengths (DSL).  We illustrate
two selection effects that occur in any diameter or/and flux
selected sample of disk galaxies: (1) there is always a Malmquist-type
selection effect that biases against disk galaxies with small disk
scale lengths at a given CSB, and (2) there could be a dependence
of the detectability of a galaxy on it disk inclination angle as 
long as disks are not completely opaque.  Without a prior knowledge 
on the full distribution function of the disk parameters,  it is
difficult to fully correct for the Malmquist bias.  On the other
hand, we derive a volume sampling function that takes into account
the inclination effect.

Using this volume sampling function, we derive a relative density
distribution of DSL for a given range of CSB values from the UGC
sample for each of the following two cases: (a) fully transparent
disks and (b) fairly opaque disks.  Replying on only the part of
the resulting distribution function that is least affected by the
Malmquist bias, we show that in the case of (a), CSB and DSL 
could be correlated in the sense that, up to an 
inclination-corrected limiting CSB of about 24.5$\,$\mpsa\ adequately
sampled by our galaxy sample, the DSL distribution of galaxies 
with a lower CSB may have a longer tail toward large values unless
the distribution of disk galaxies as a function of CSB rises very
rapidly toward faint values.  In the case of (b), the face-on limiting 
CSB of our galaxy sample is too faint to set a useful constraint
on the faint part of the CSB distribution function.

\acknowledgments

The author is grateful to E. E. Salpeter and G. L. Hoffman for their
comments on this paper and contribution to the early part of our CCD
surface photometry campaign; to the editor, G. Bothun, and the referee,
S. McGaugh, for a number of comments that
helped improving this paper;  to W. Freudling for providing us with
the GALPHOT software package which is part of the data reduction 
tools used in this paper; to M. Schmitz for going over the tables; 
and to the staff members of Palomar Observatory for their assistance
during the observation.  The observations 
presented in this paper were made at the Palomar Observatory as 
part of a continuing collaborative agreement between the California
Institute of Technology and Cornell University.   This work was 
supported in part by Jet Propulsion Laboratory,  California Institute
of Technology, under a contract with  the National Aeronautics and 
Space Administration.
\vspace{5pt}


\newpage

\addtocounter{page}{-5}
\begin{figure}
\plotone{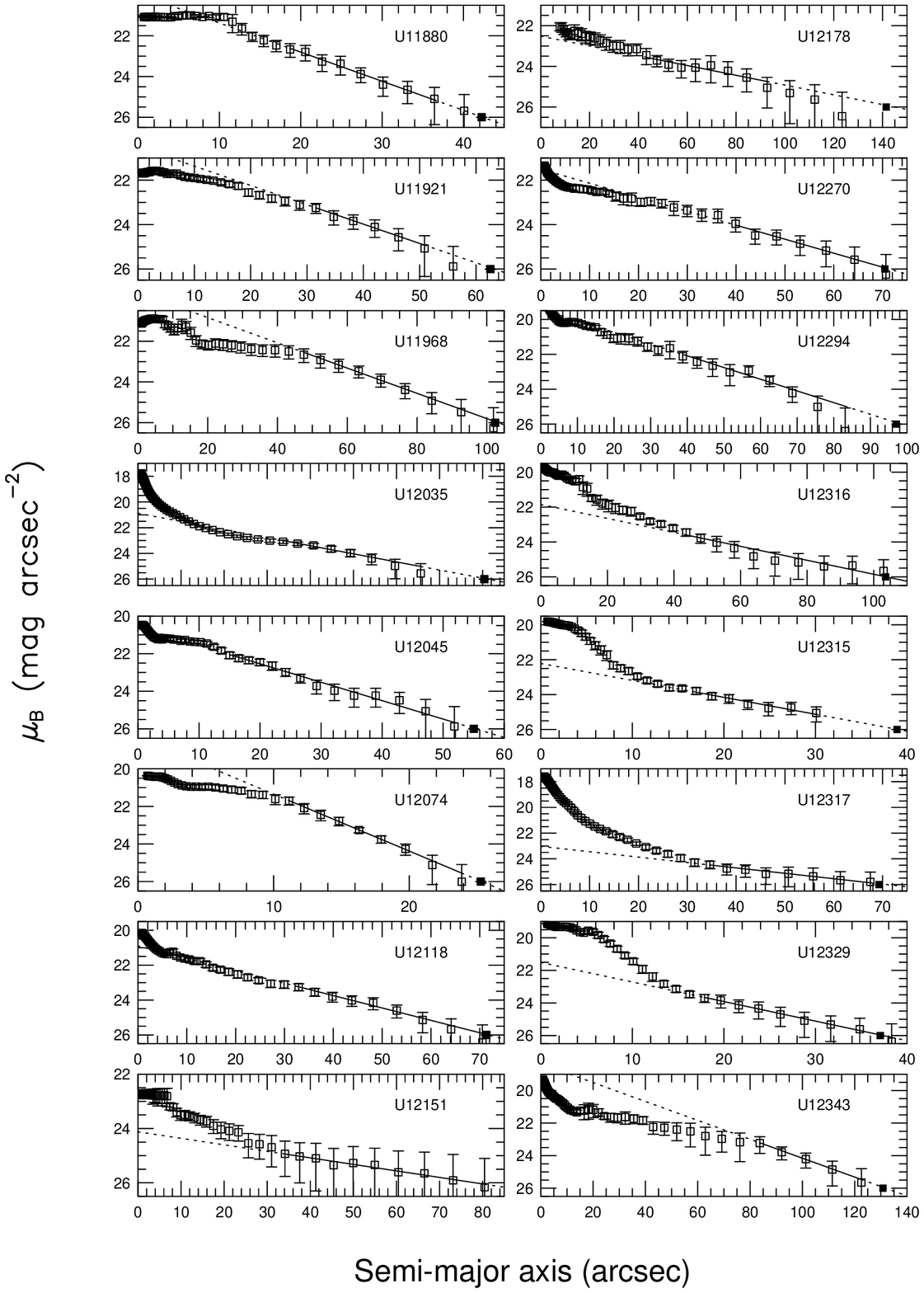}
\vspace{-0.4in}
\caption{
     Plots of the B-band surface brightness as a function of the semi-major
     axis for the galaxies with photometric data in Table~1.  The error bars
     shown are r.m.s.~statistical errors.   Open squares are measured 
     isophotes. The solid-to-dashed line is the exponential disk from 
     a fit to the data points covered by the solid line segment.  The filled
     square indicates the B 26th\mpsa\ isophote determined from 
     the exponential disk fit.}
\end{figure}
\setcounter{figure}{0}
\begin{figure}
\plotone{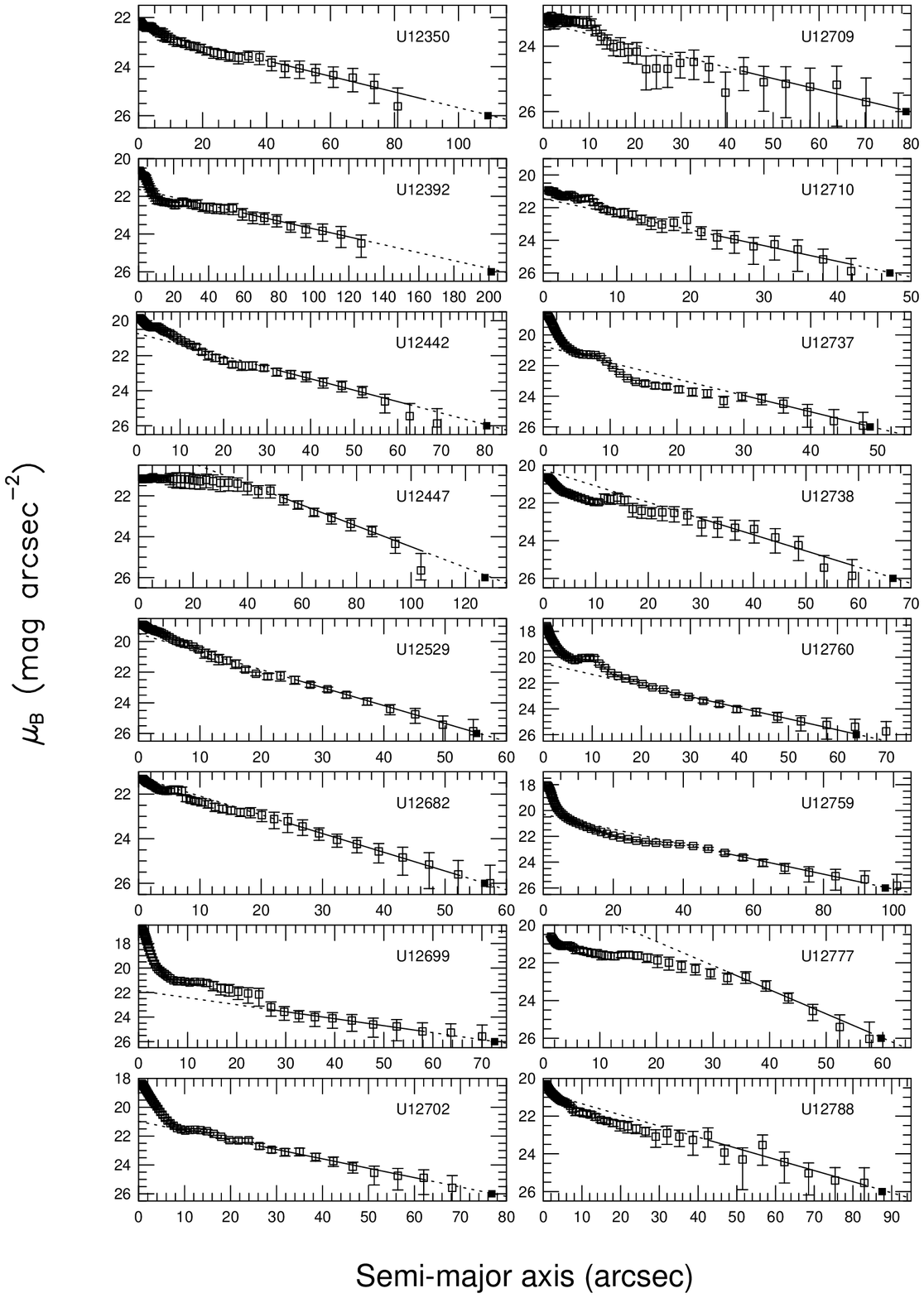}
\vspace{-0.4in}
\caption{Continued.}
\end{figure}
\setcounter{figure}{0}
\begin{figure}
\plotone{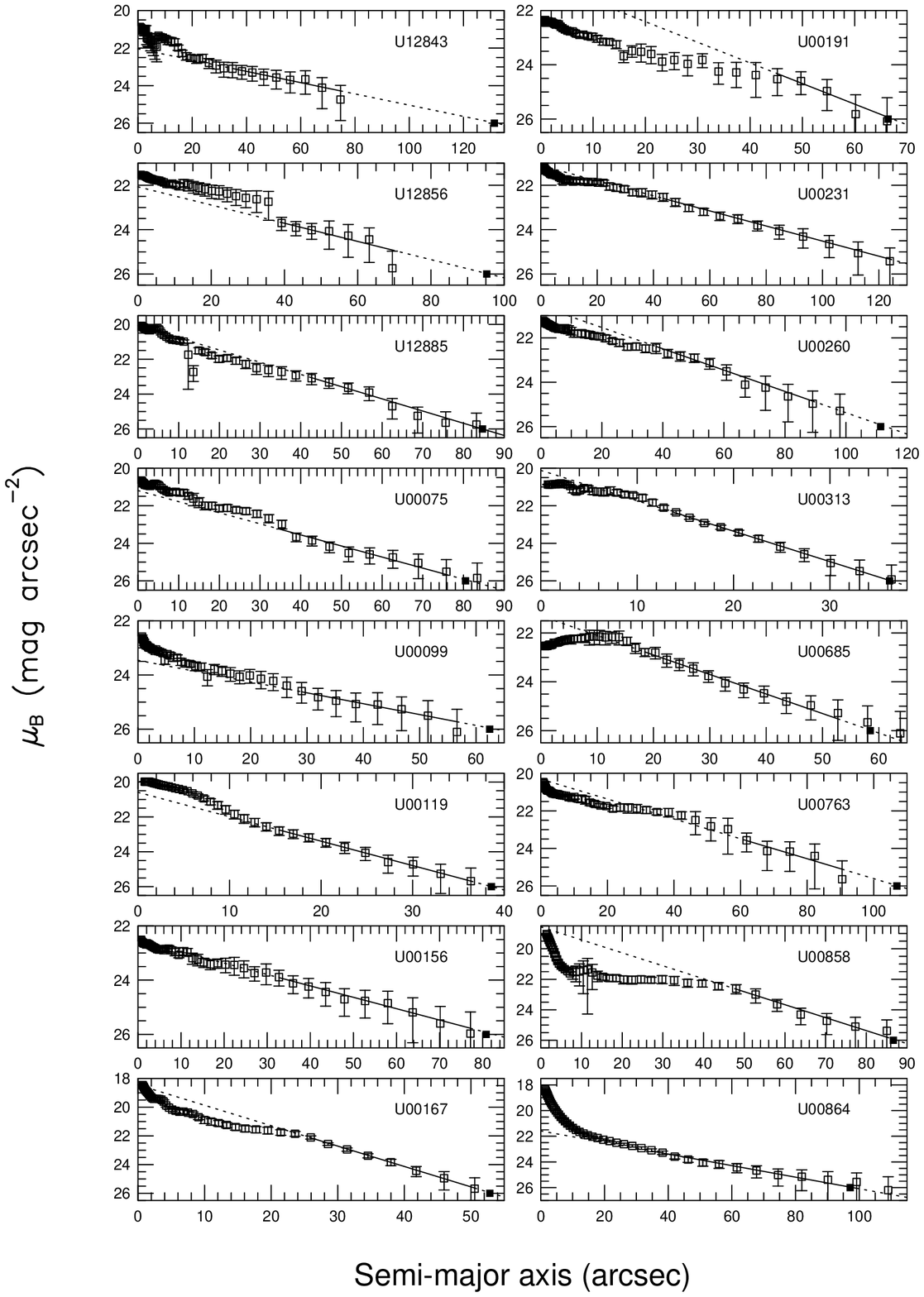}
\vspace{-0.4in}
\caption{Continued.}
\end{figure}

\setcounter{figure}{0}
\begin{figure}
\plotone{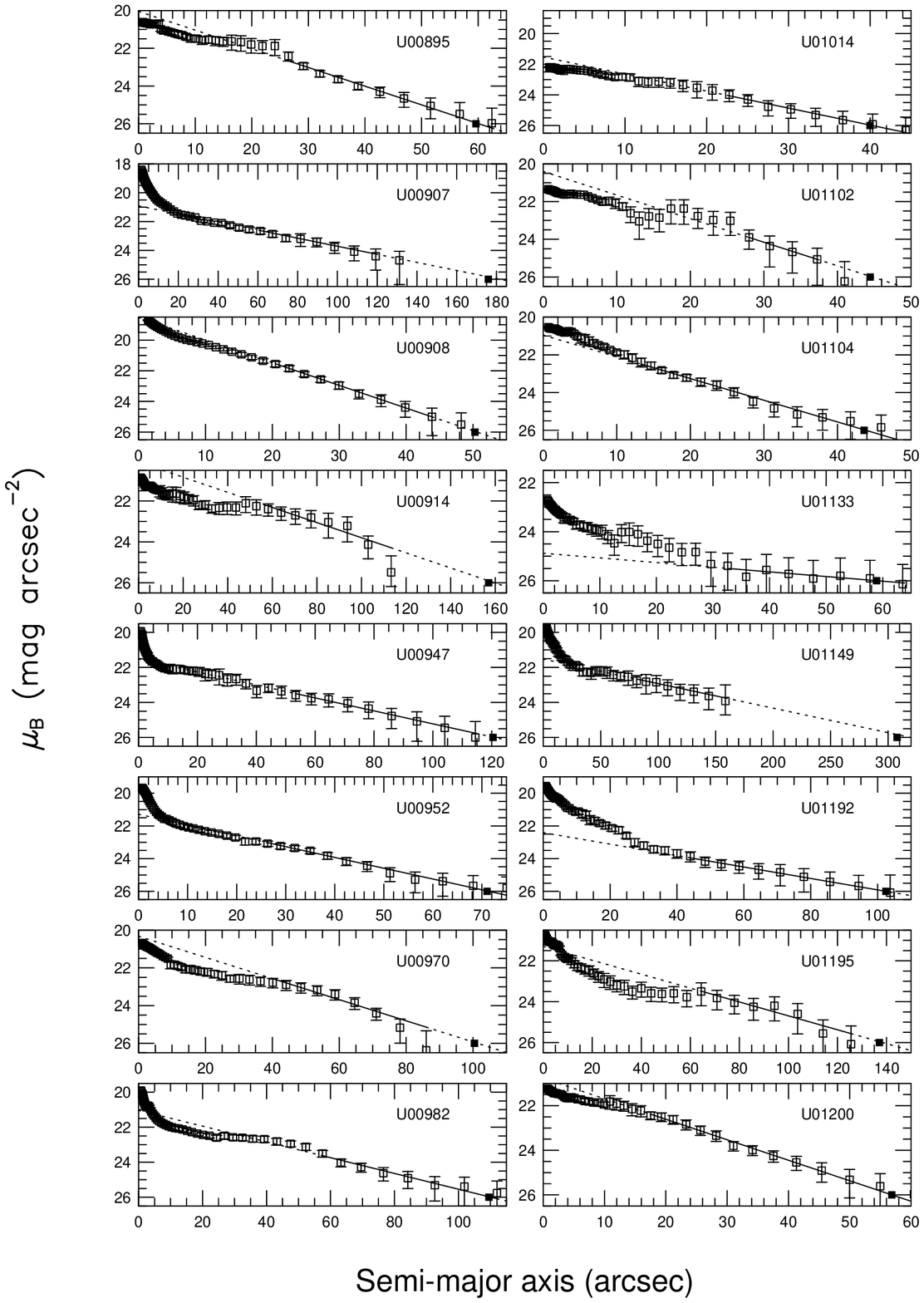}
\vspace{-0.4in}
\caption{Continued.}
\end{figure}

\setcounter{figure}{0}
\begin{figure}
\plotone{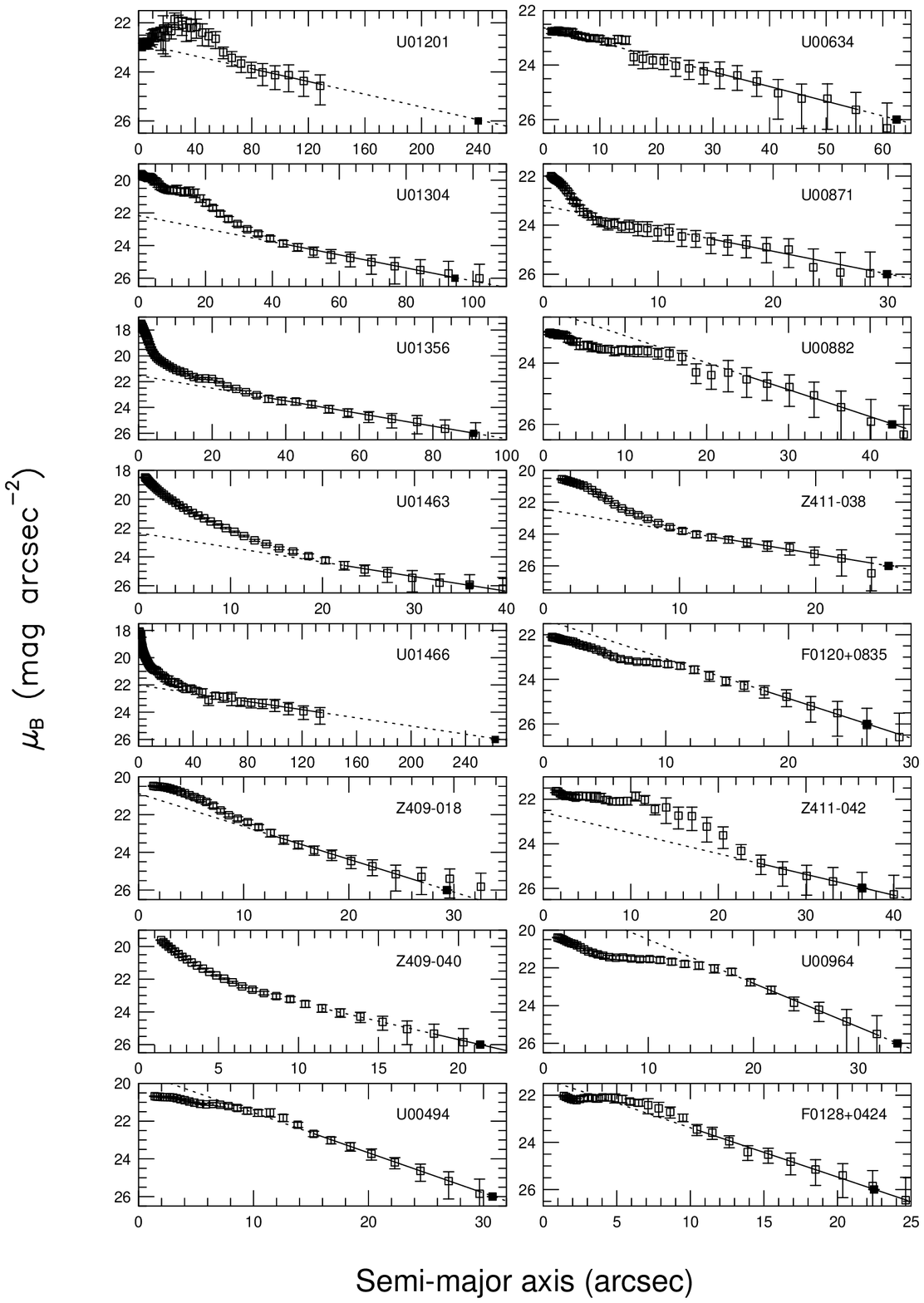}
\vspace{-0.4in}
\caption{Continued.}
\end{figure}

\addtocounter{page}{5}
\setcounter{figure}{1}

\begin{figure}
\plotone{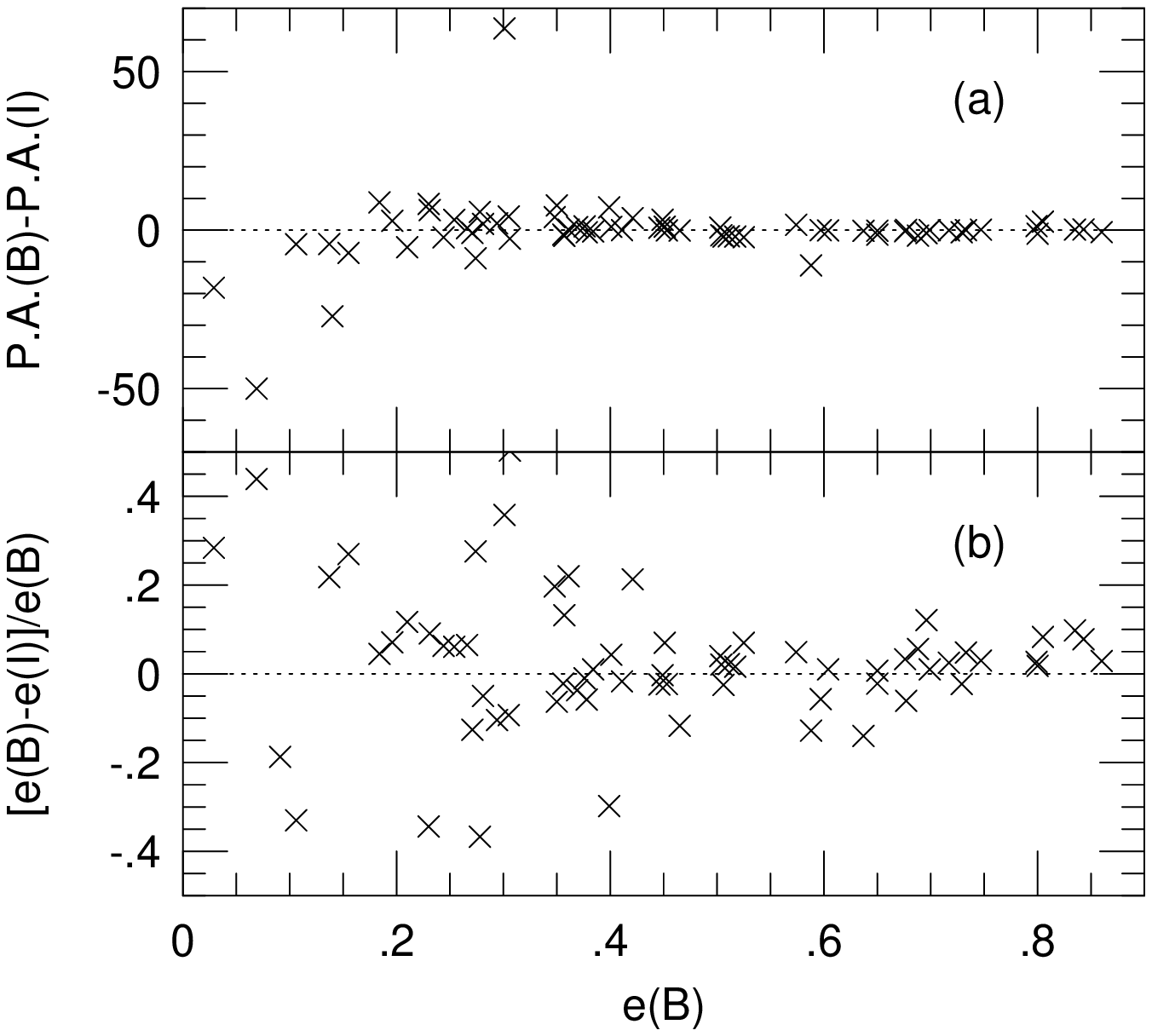}
\vspace{-1.5in}
\caption{
     Plots as a function of the B-band disk ellipticity of (a) the difference
     in the mean disk position angle and (b) the relative difference in 
     the mean disk ellipticity between the B-band result in this paper 
     and the I-band result in Lu \etal (1993).  Only UGC sample galaxies
     with available B and I data are shown here.}
\end{figure}

\begin{figure}
\plotone{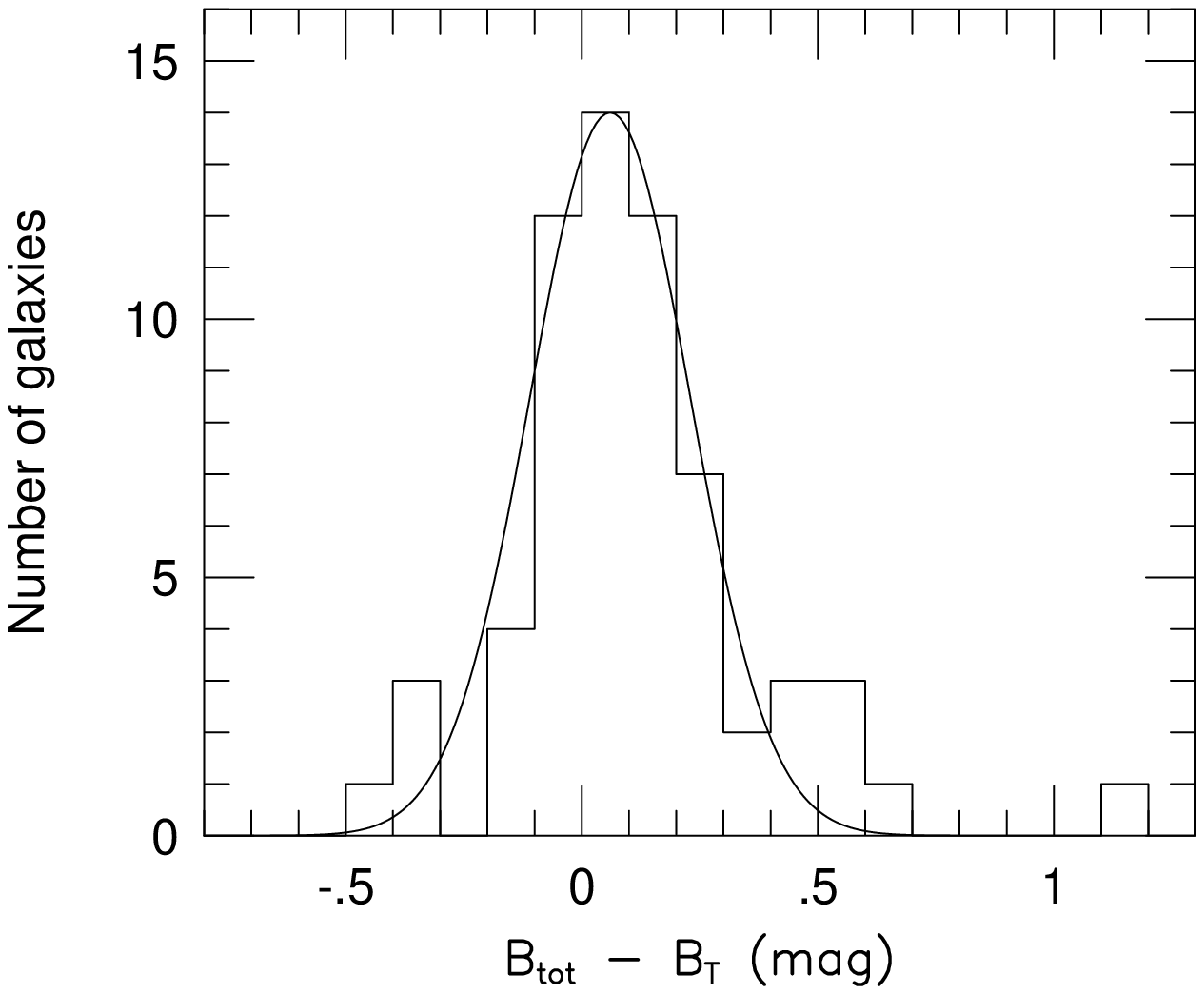}
\vspace{-2.0in}
\caption{
     Histogram distribution of ($B\tot - B_T$) for our UGC sample, where
     $B_T$ is taken from the following sources arranged in decreasing 
     preference:  $B_T$ in RC3, $m_B$ in RC3, and the total B magnitude
     estimated in Lu \etal (1993).  None of the galaxies with notes in 
     Table~1 are used here.  The largest magnitude offset in the figure
     is from UGC~00099, a galaxy of Sd/Sm type without available $B_T$
     or $m_B$ in RC3.  The Gaussian curve shown centers at ($B\tot - I\tot$)
     $= 0.06$ mag and has a full width at half maximum of $0.4$ mag.}
\end{figure}

\begin{figure}
\plotone{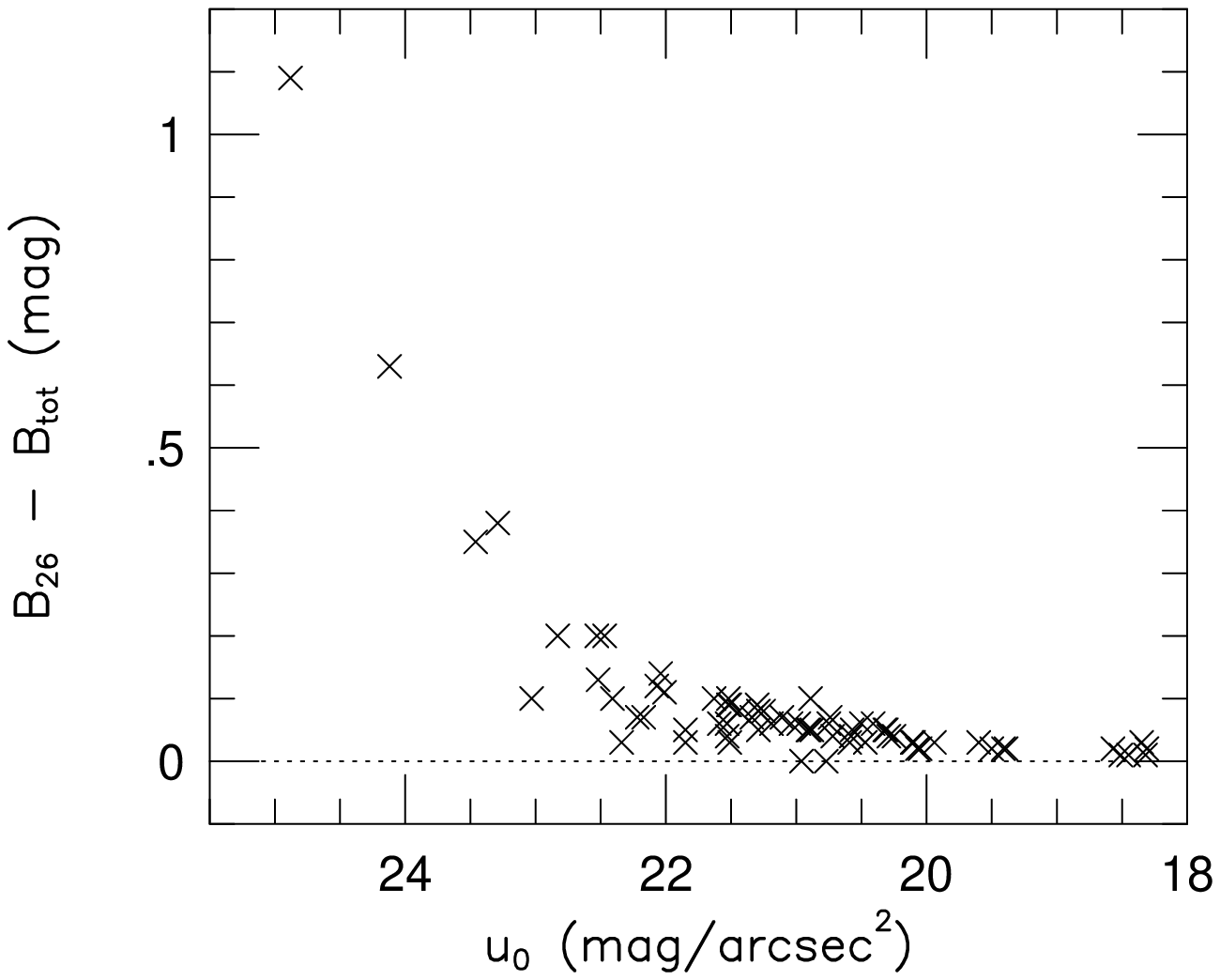}
\vspace{-1.0in}
\caption{
     Plot of $(B_{26}-B\tot)$ as a function of the central disk surface 
     brightness for the UGC sample galaxies with available photometry in 
     Table~1, where $B_{26}$ is the isophotal magnitude at B 26th-\mpsa\
     and $B\tot$ is the total magnitude.}
\end{figure}

\begin{figure}
\plotone{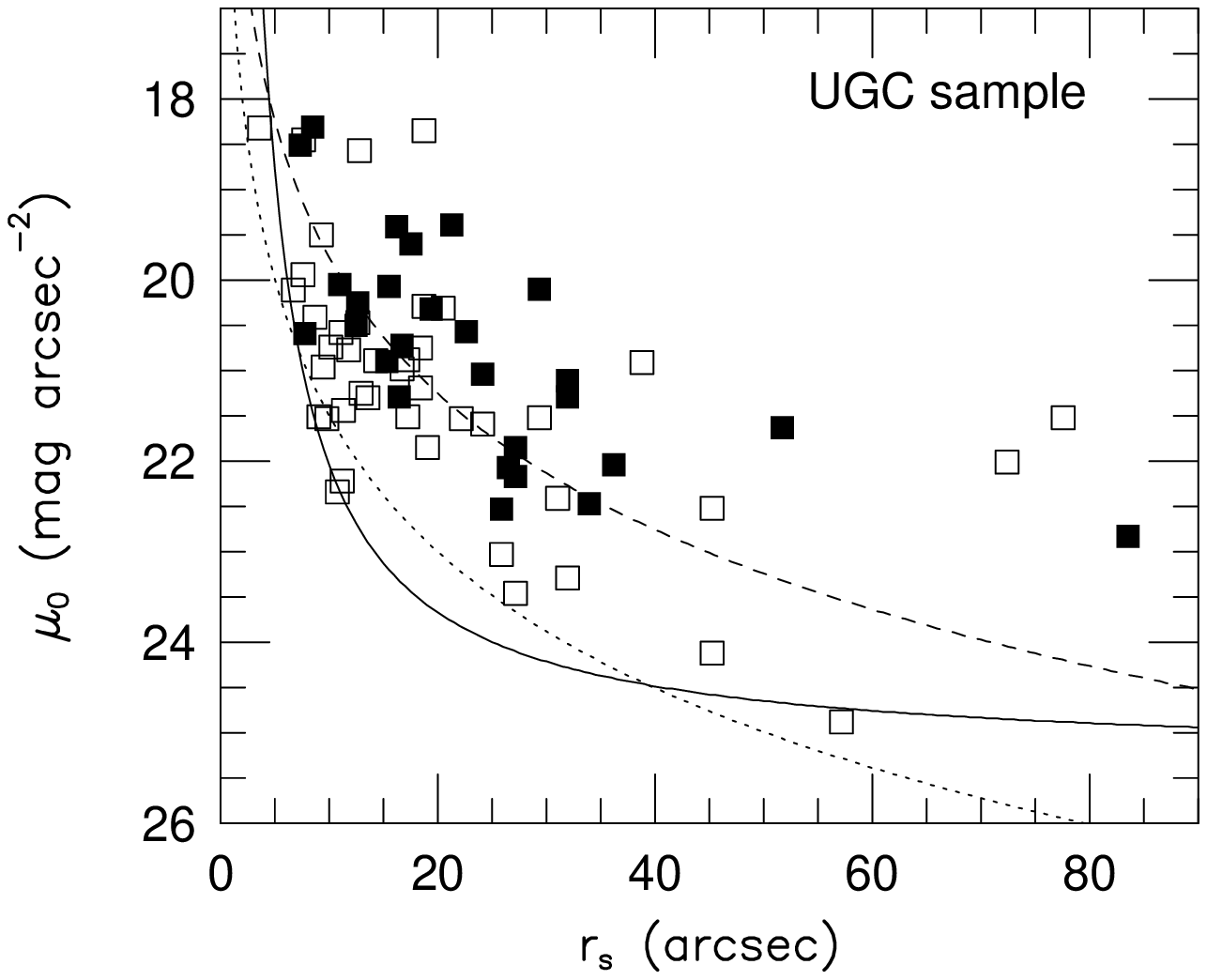}
\vspace{-2.0in}
\caption{
     Plot of the observed central surface brightness as a function of
     the observed angular exponential scale length for the disks of 
     the galaxies in the UGC sample.  The solid curve represents 
     the diameter selection, eq.~(3) in the text; while the dotted 
     and dashed curves sketch respectively the magnitude selection,
     eq.~(5) in the text, for the cases of $e = 0$ and $e = 0.8$,
     respectively.  The filled (open) squares are galaxies with 
     ellipticities greater (less) than $0.5$.}
\end{figure}

\begin{figure}
\plotone{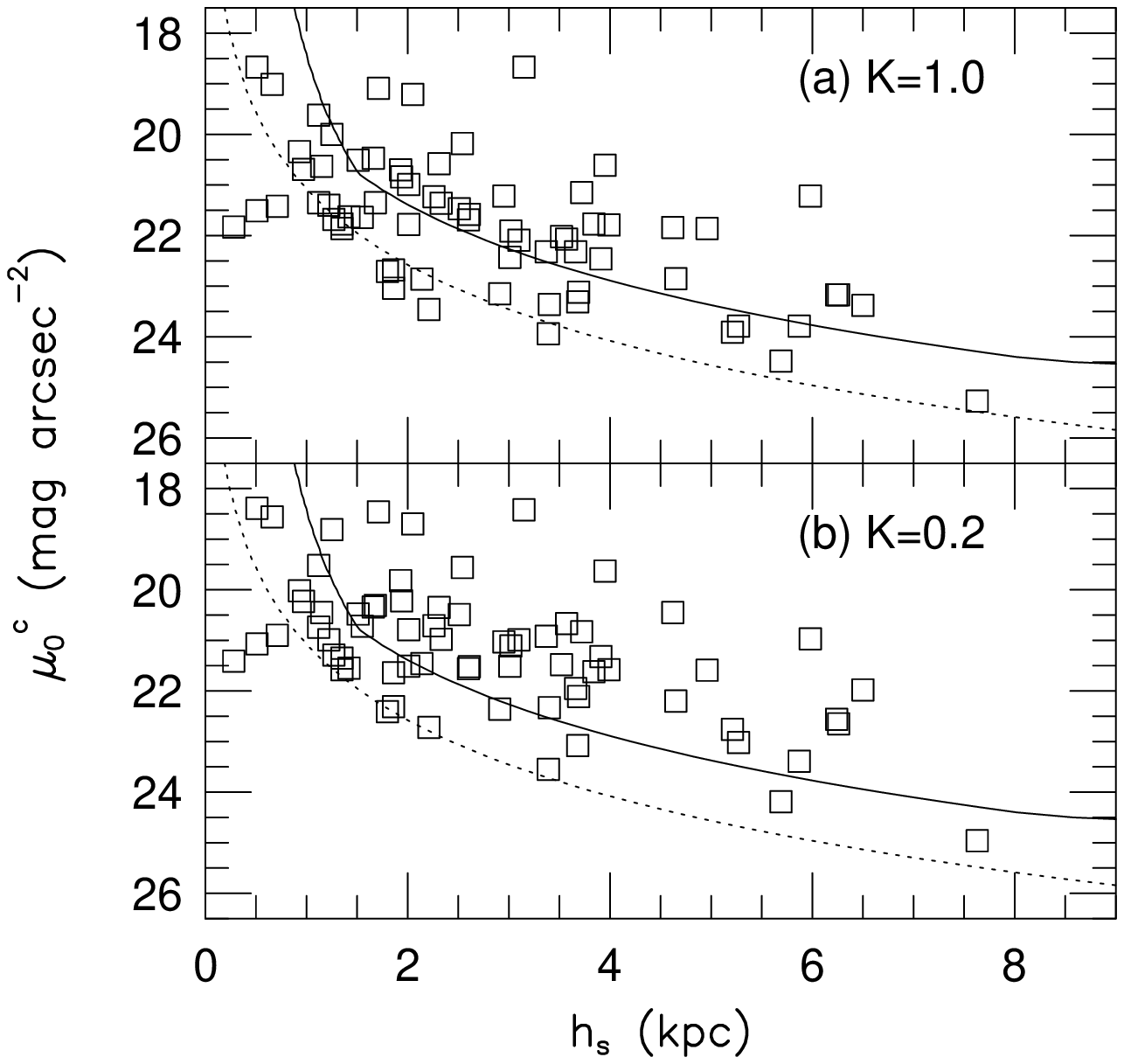}
\vspace{-2.0in}
\caption{
     Plot of the face-on central surface brightness
     as a function of the linear disk scale length for the UGC sample
     galaxies.  The dotted curve, given by columns~(4) Table~2, 
     indicates the path of a face-on disk galaxy of $M_B = -17.5\,$mag
     which is roughly detectable up to half of the maximum UGC sample
     distance.  The solid curve, given by column~(5) of Table~2, 
     indicates the threshold above which (and to whose right) a 
     face-on galaxy will be detectable up to the maximum sample distance.}
\end{figure}

\begin{figure}
\plotone{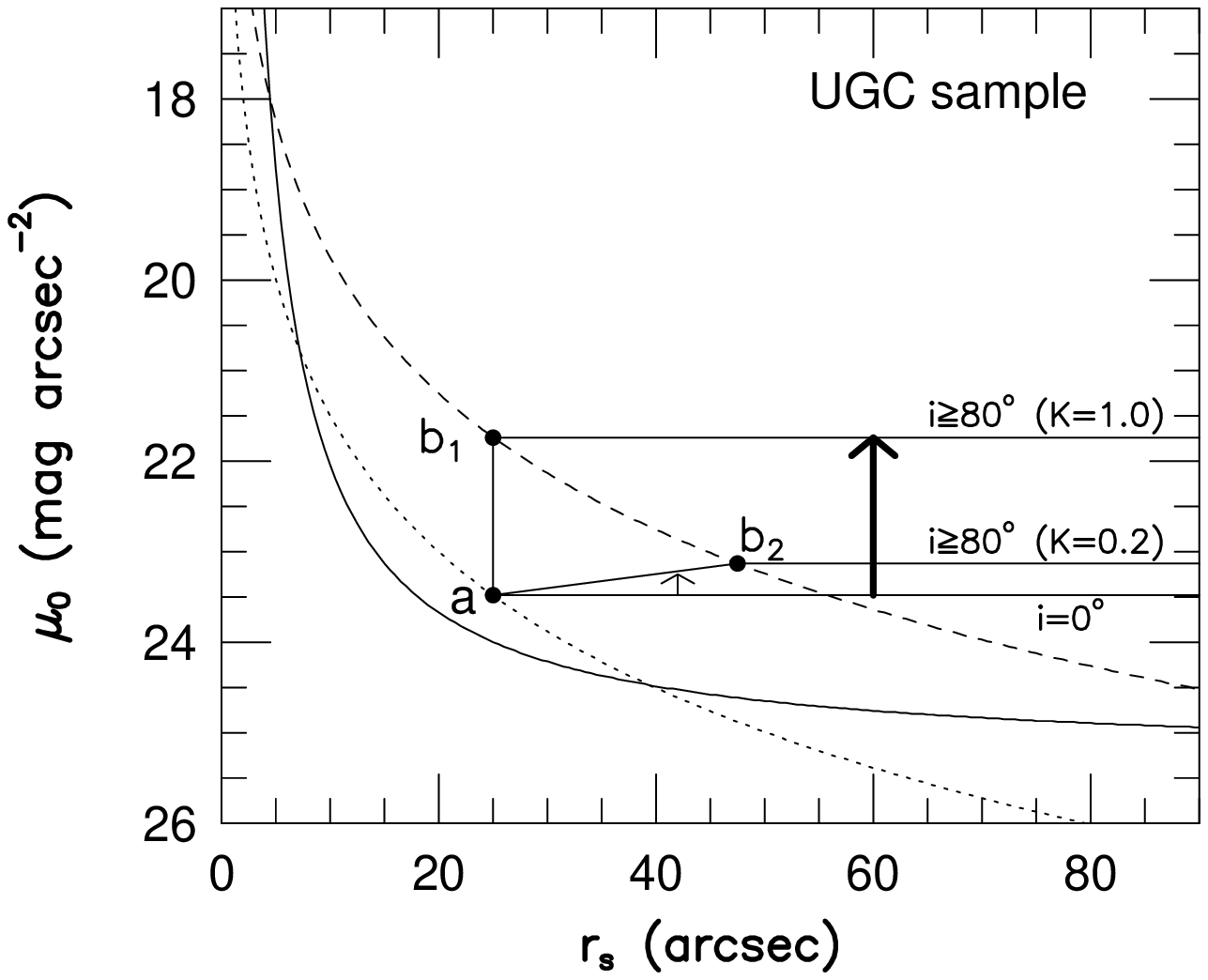}
\vspace{-2.0in}
\caption{
     A sketch illustrating how a phase space for a disk galaxy is 
     constructed in terms of (1) real space defined by $\mu_0$ 
     and $r_s$, and (2) a phase space defined by disk inclination
     angle through eq.~(8) in the text.   The curves are the same
     as in Fig.~5.  The horizontal lines indicate a disk inclination
     of $i=0\md$ and $i\ge 80\md$, respectively.  The thin (thick)
     arrow sketches how a fully (partially) transparent disk moves
     as its disk inclination increases at a given distance.}
\end{figure}

\begin{figure}
\plotone{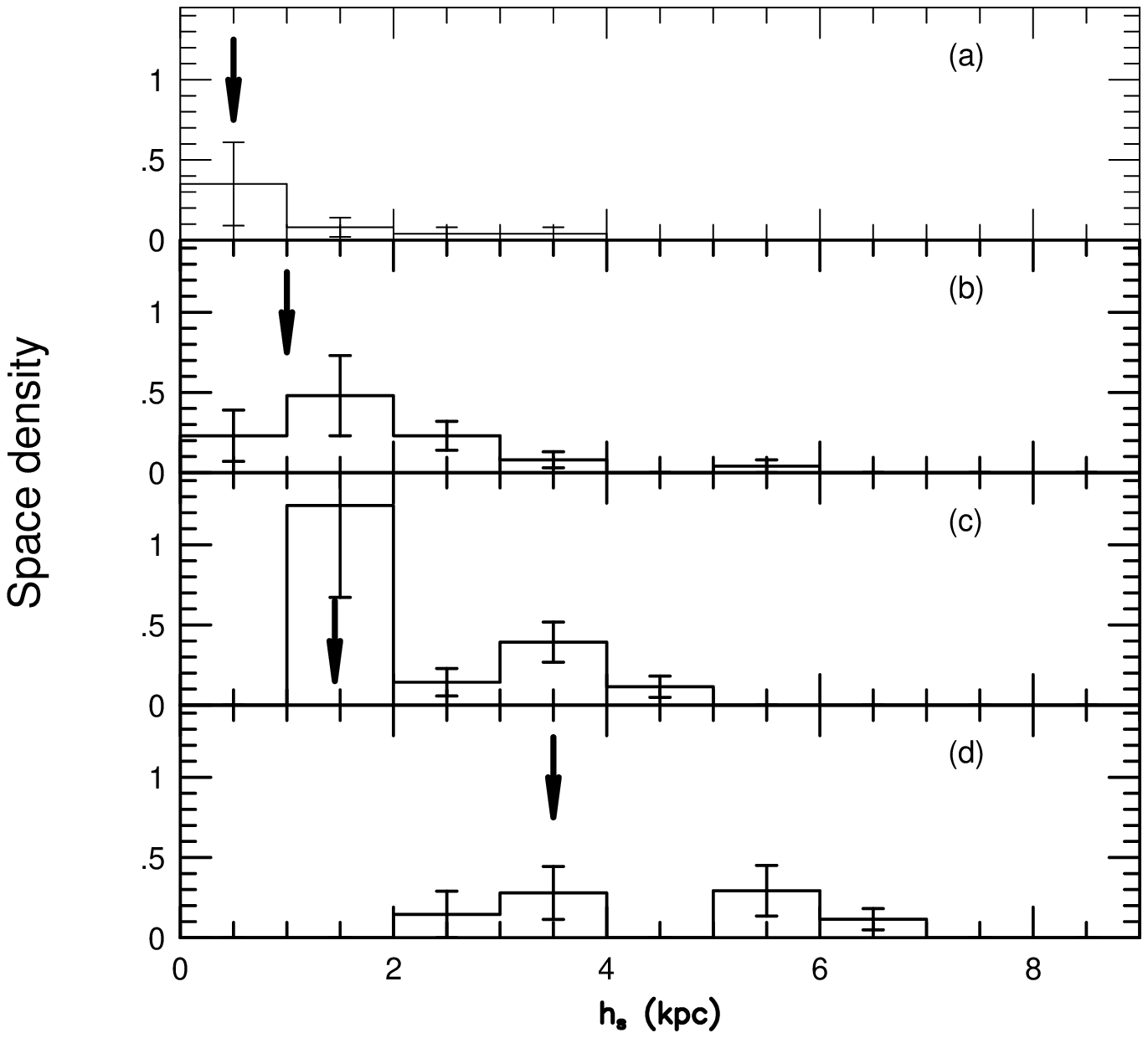}
\vspace{-2.0in}
\caption{
     Plots of the volume sampling function-adjusted distribution of 
     the disk scale lengths for UGC sample galaxies in that case of 
     transparent disks with $K = 1$: (a) $18 < \mu_0^c < 20$\mpsa,  
     (b) $20 < \mu_0^c < 21.5$\mpsa, (c) $21.5 < \mu_0^c < 23$\mpsa, and
     (d) $\mu_0^c > 23$\mpsa.  The ordinate is in units of number of
     galaxies per $10^3\Omega\,$Mpc$^3$, where $\Omega$ ($\approx 0.13\,$sr)
     is the solid angle covered by our UGC sample on the sky.
     Only sample galaxies with $M_B < -17.5\,$mag are used.  The error 
     bars are weighted r.m.s.~values assuming Poisson statistics.}
\end{figure}

\begin{figure}
\plotone{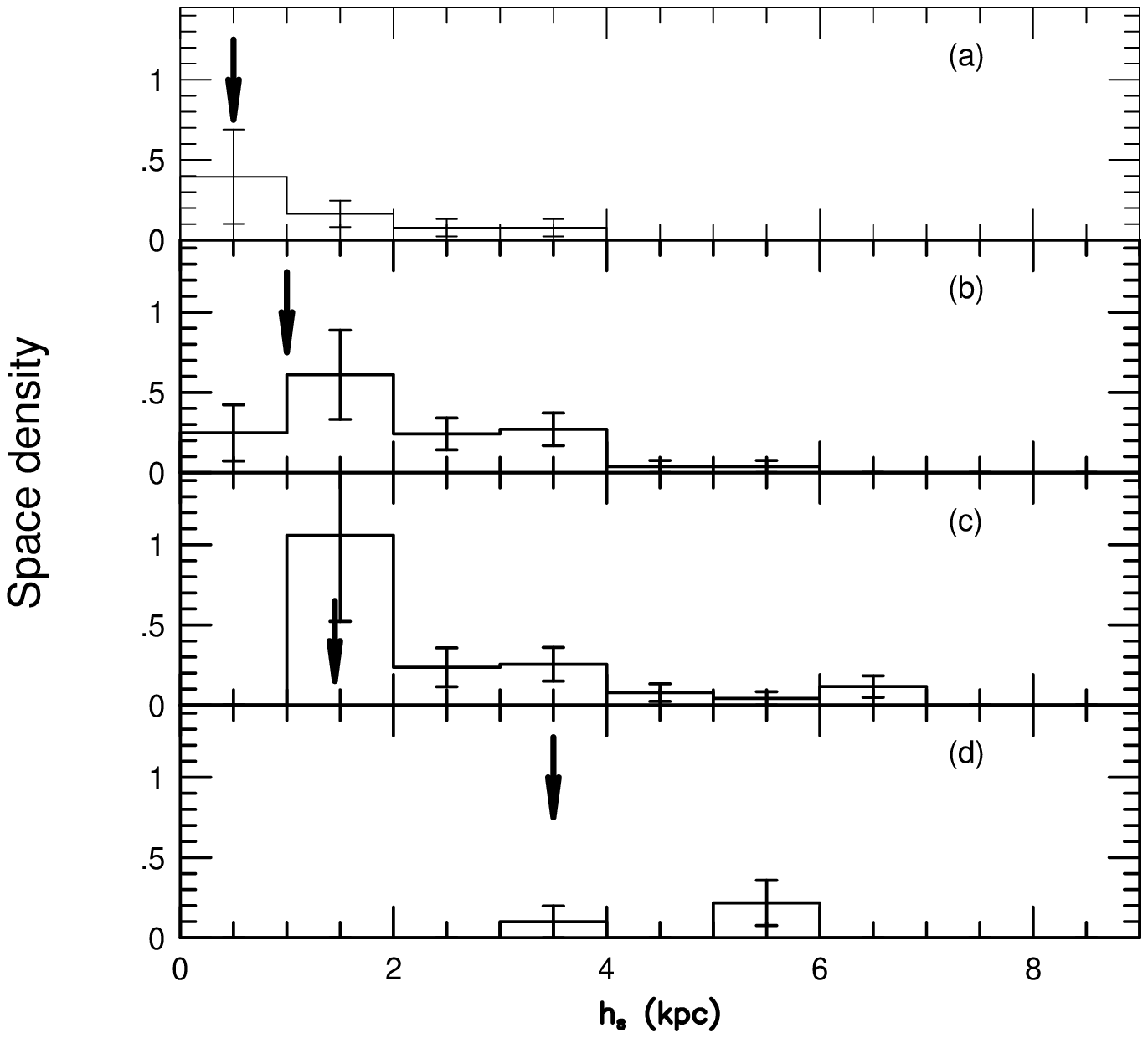}
\vspace{-0.5in}
\caption{
     The same as in Fig.~8, but for the case of fairly opaque disks 
     with $K = 0.2$.}
\end{figure}
\newpage

\newcommand{\no}{\nodata}
\begin{deluxetable}{llrrrcrrrrrrcccc}
\scriptsize
\tablenum{1}
\tablewidth{0pt}
\tablecaption{B-band CCD Surface Photometric Results}
\tablehead{
\colhead{UGC}   & \colhead{NGC/IC}  & \colhead{Dist.}  & \colhead{$r_1$}     & \colhead{$r_2$}    &
\colhead{$e$}   & \colhead{$i$}   & \colhead{P.A.}    & \colhead{$B_{26}$}  & \colhead{$D_{26}$} &
\colhead{$r_s$} & \colhead{$h_s$} & \colhead{$\mu_0$} & \colhead{$B_{tot}$} & \colhead{$M_B$} & \colhead{$(B-I)$} \\
\colhead{(1)}   & \colhead{(2)}  & \colhead{(3)}   & \colhead{(4)}  &
\colhead{(5)}   & \colhead{(6)}  & \colhead{(7)}   & \colhead{(8)}  &
\colhead{(9)}   & \colhead{(10)}  & \colhead{(11)} & \colhead{(12)}  &
\colhead{(13)}  & \colhead{(14)}  & \colhead{(15)} & \colhead{(16)}}

\startdata
I. \ The UGC & Sample:\nl
U11880\phn & I1420\phn  & 25.33 & 18.7 &  36.4 & 0.306 & 47 & 112 & 14.20 & 1.41 &  7.6 &  0.93  &19.94 & 14.17&  -17.85 & 1.54 \nl
U11921     & \no        & 25.63 & 31.6 &  50.9 & 0.650 & 72 & 124 & 14.62 & 2.09 & 12.5 &  1.55  &20.50 & 14.56&  -17.48 & 1.37 \nl
U11968     & N7241      & 22.71 & 47.6 & 101.9 & 0.637 & 71 &  21 & 13.02 & 3.41 & 17.5 &  1.93  &19.60 & 12.99&  -18.79 & 1.94 \nl
U12035     & N7280      & 27.91 & 33.7 &  65.6 & 0.356 & 51 &  74 & 13.09 & 2.68 & 17.2 &  2.33  &20.88 & 13.04&  -19.19 & 2.10 \nl
U12045     & N7290      & 42.04 & 15.0 &  51.8 & 0.449 & 58 & 161 & 14.03 & 1.83 & 11.1 &  2.26  &20.58 & 13.99&  -19.13 & 1.58 \nl
\nl											 
U12074     & \no        & 29.42 & 11.1 &  23.9 & 0.281 & 45 & 149 & 14.71 & 0.84 &  3.6 &  0.51 & 18.32 & 14.70 & -17.64 & 2.04 \nl
U12118     & N7328      & 40.67 & 32.9 &  70.5 & 0.604 & 69 &  86 & 13.95 & 2.38 & 15.3 &  3.02 & 20.90 & 13.90 & -19.15 & 1.96 \nl
U12151     & \no        & 25.94 & 34.1 &  80.5 & 0.278 & 44 &   0 & 14.80 & 2.61 & 45.2 &  5.69 & 24.12 & 14.17 & -17.90 & 1.36 \nl
U12178\tablenotemark{a} & \no  								 
			& 28.54 & 43.3 &  92.7 & 0.453 & 58 &   9 & 13.06 & 4.72 & 45.2 &  6.26 & 22.52 & 12.87 & -19.75 & 1.24 \nl
U12270     & N7437      & 31.28 & 39.9 &  70.7 & 0.069 & 21 &  43 & 13.57 & 2.35 & 17.2 &  2.61 & 21.51 & 13.48 & -19.00 & 1.48 \nl
\nl	
U12294     & N7448      & 32.37 & 38.8 &  83.1 & 0.511 & 62 & 171 & 12.24 & 3.24 & 16.2 &  2.54 & 19.41 & 12.22 & -20.33 & 1.48 \nl
U12316     & N7463      & 28.02 & 43.7 & 113.2 & 0.688 & 75 &  93 & 13.37 & 3.45 & 27.1 &  3.69 & 21.85 & 13.34 & -18.90 & 1.42 \nl
U12315     & N7464      & 34.26 & 17.0 &  30.1 & 0.348 & 50 & 154 & 14.71 & 1.30 & 11.2 &  1.86 & 22.22 & 14.64 & -18.03 & 1.01 \nl
U12317     & N7465      & 29.35 & 34.6 &  67.5 & 0.230 & 40 & 165 & 13.30 & 2.31 & 25.9 &  3.68 & 23.03 & 13.20 & -19.14 & 1.72 \nl
U12329     & N7468      & 30.89 & 17.9 &  38.3 & 0.274 & 44 &  19 & 14.19 & 1.24 &  9.1 &  1.35 & 21.51 & 14.16 & -18.29 & 1.30 \nl
\nl		               	 		        				          		           			       	
U12343     & N7479      & 34.71 & 83.7 & 122.6 & 0.254 & 42 &  39 & 11.81 & 4.36 & 18.7 &  3.15 & 18.35 & 11.78 & -20.92 & 1.99 \nl
U12350     & \no        & 31.66 & 41.6 &  89.1 & 0.733 & 79 &  95 & 14.48 & 3.64 & 33.9 &  5.21 & 22.47 & 14.28 & -18.22 & 1.52 \nl
U12392     & N7497      & 25.93 & 59.3 & 127.1 & 0.800 & 90 &  42 & 12.98 & 6.71 & 51.7 &  6.50 & 21.63 & 12.88 & -19.19 & 1.87 \nl
U12442     & N7537      & 38.25 & 32.2 &  62.8 & 0.717 & 78 &  77 & 13.85 & 2.68 & 16.7 &  3.10 & 20.72 & 13.81 & -19.10 & 1.81 \nl
U12447     & N7541      & 38.23 & 53.2 & 103.6 & 0.676 & 74 & 100 & 12.37 & 4.24 & 21.3 &  3.95 & 19.39 & 12.35 & -20.56 & 2.06 \nl
\nl	
U12529     & N7625      & 24.83 & 25.5 &  54.6 & 0.106 & 27 &  32 & 12.86 & 1.84 &  9.3 &  1.12 & 19.50 & 12.84 & -19.13 & 2.02 \nl
U12613\tablenotemark{b} & \no  
                        & \no   & \no  &  \no  & \no   & \no & \no & \no  &  \no &  \no &  \no  &  \no  &  \no  & \no    & \no  \nl
U12682     & \no        & 21.57 & 24.3 &  52.1 & 0.378 & 53 &  43 & 14.35 & 1.88 & 12.9 &  1.35 & 21.25 & 14.27 & -17.40 & 1.27 \nl
U12699     & N7714      & 39.59 & 27.0 &  57.9 & 0.350 & 50 &  37 & 13.04 & 2.42 & 19.1 &  2.66 & 21.85 & 12.99 & -20.00 & 1.69 \nl
U12702     & N7716      & 36.48 & 26.3 &  62.0 & 0.196 & 37 &  45 & 13.03 & 2.56 & 16.7 &  2.95 & 20.98 & 12.97 & -19.84 & 1.85 \nl
\nl	
U12709     & \no        & 37.88 & 43.6 &  77.2 & 0.369 & 52 & 144 & 14.69 & 2.63 & 31.9 &  5.87 & 23.29 & 14.31 & -18.58 & 1.44 \nl
U12710     & \no        & 36.58 & 23.6 &  41.9 & 0.266 & 43 & 162 & 14.48 & 1.57 & 11.3 &  2.01 & 21.44 & 14.40 & -18.42 & 1.20 \nl
U12737     & N7731      & 40.81 & 29.7 &  47.8 & 0.210 & 38 &  91 & 14.26 & 1.63 & 10.1 &  2.01 & 20.74 & 14.19 & -18.86 & 1.88 \nl
U12738     & N7732      & 41.09 & 30.2 &  58.8 & 0.677 & 74 &  89 & 14.33 & 2.22 & 12.6 &  2.51 & 20.25 & 14.29 & -18.78 & 1.51 \nl
U12760     & N7742      & 24.65 & 24.5 &  63.6 & 0.029 & 14 &  50 & 12.50 & 2.13 & 12.6 &  1.51 & 20.47 & 12.47 & -19.49 & 1.87 \nl
\nl	
U12759     & N7743      & 25.40 & 51.8 &  91.8 & 0.231 & 40 &  75 & 12.42 & 3.25 & 18.7 &  2.31 & 20.29 & 12.38 & -19.64 & 2.16 \nl
U12777     & N7750      & 41.47 & 32.5 &  57.6 & 0.517 & 63 & 173 & 13.56 & 1.99 &  8.5 &  1.71 & 18.31 & 13.54 & -19.55 & 1.77 \nl
U12788     & N7757      & 41.71 & 42.6 &  82.9 & 0.305 & 47 & 123 & 13.09 & 2.92 & 18.4 &  3.72 & 20.75 & 13.03 & -20.07 & 1.26 \nl
U12843\tablenotemark{c} & \no       
			& 26.53 & 34.8 &  74.7 & 0.525 & 63 &  24 & 13.14 & 4.38 & 36.2 &  4.65 & 22.04 & 13.00 & -19.12 & 1.12 \nl
U12856     & \no        & 26.45 & 39.2 &  69.5 & 0.696 & 76 &  15 & 14.15 & 3.17 & 26.5 &  3.40 & 22.07 & 14.03 & -18.08 & 0.96 \nl
\nl	
U12885     & N7800      & 25.84 & 42.7 &  91.5 & 0.506 & 62 &  45 & 13.28 & 2.82 & 15.5 &  1.94 & 20.07 & 13.26 & -18.80 & 1.36 \nl
U00008\tablenotemark{d} & N7814     
			& \no   & \no  &  \no  & \no   &\no & \no & \no   & \no  & \no  &  \no  & \no   & \no   & \no    & \no  \nl
U00075     & N0014      & 14.19 & 35.4 &  75.8 & 0.361 & 51 &  28 & 13.28 & 2.68 & 18.4 &  1.27 & 21.19 & 13.22 & -17.54 & 1.69 \nl
U00099     & \no        & 25.77 & 29.1 &  56.6 & 0.357 & 51 & 154 & 15.00 & 2.08 & 27.1 &  3.39 & 23.46 & 14.65 & -17.41 & 1.24 \nl
U00119     & \no        & 29.66 & 15.4 &  36.4 & 0.504 & 62 &  81 & 14.62 & 1.29 &  7.8 &  1.12 & 20.59 & 14.59 & -17.77 & 1.34 \nl
\nl	
U00156     & \no        & 17.60 & 32.7 &  77.2 & 0.574 & 67 &   0 & 14.62 & 2.69 & 25.9 &  2.21 & 22.53 & 14.42 & -16.81 & 1.58 \nl
U00167     & N0063      & 17.87 & 25.9 &  50.6 & 0.401 & 54 & 104 & 13.17 & 1.76 &  7.6 &  0.66 & 18.45 & 13.16 & -18.10 & 1.89 \nl
U00191     & \no        & 17.61 & 45.2 &  66.2 & 0.376 & 52 & 154 & 14.42 & 2.21 & 14.3 &  1.22 & 20.89 & 14.32 & -16.91 & 1.60 \nl
U00231     & N0100      & 13.82 & 43.4 & 123.9 & 0.860 & 90 &  55 & 13.84 & 4.66 & 31.9 &  2.14 & 21.11 & 13.77 & -16.93 & 1.71 \nl
U00260     & \no        & 30.76 & 37.8 &  89.1 & 0.843 & 90 &  22 & 14.03 & 3.71 & 22.6 &  3.37 & 20.57 & 13.98 & -18.46 & 1.57 \nl
\nl	
U00313     & \no        & 29.84 & 14.0 &  36.4 & 0.411 & 55 &  13 & 14.62 & 1.21 &  6.7 &  0.97 & 20.11 & 14.59 & -17.78 & 1.61 \nl
U00668\tablenotemark{b} & I1613 
                        & \no   & \no  &  \no  &  \no  &\no & \no &  \no  &  \no &  \no &  \no  & \no   & \no   & \no    & \no  \nl
U00685     & \no        &  4.27 & 18.5 &  52.7 & 0.384 & 53 & 116 & 14.32 & 1.95 & 13.6 &  0.28 & 21.30 & 14.23 & -13.92 & 1.51 \nl
U00711\tablenotemark{d}  & \no 
                        &  \no  &  \no & \no   & \no   &\no & \no &  \no  &  \no &  \no &  \no  & \no   &  \no  &   \no  & \no  \nl
U00763     & N0428      & 16.76 & 61.8 &  90.5 & 0.140 & 31 &  73 & 12.17 & 3.57 & 20.5 &  1.66 & 20.31 & 12.12 & -19.00 & 1.43 \nl
\nl	
U00858     & N0470      & 33.04 & 48.0 &  85.0 & 0.446 & 58 & 151 & 12.69 & 2.89 & 12.8 &  2.05 & 18.57 & 12.67 & -19.93 & 1.98 \nl
U00859\tablenotemark{e}  & N0473 
                        &  \no  &  \no & \no   & \no   &\no & \no &  \no  &  \no &  \no &  \no  & \no   &  \no  &   \no  & \no  \nl
U00864     & N0474      & 32.85 & 34.8 &  99.2 & 0.155 & 33 &  10 & 12.62 & 3.24 & 24.1 &  3.84 & 21.59 & 12.56 & -20.02 & 2.35 \nl
U00895     & N0485      & 31.56 & 26.5 &  62.4 & 0.699 & 76 &   4 & 14.24 & 1.99 & 11.0 &  1.68 & 20.05 & 14.22 & -18.28 & 1.91 \nl
U00907     & N0488      & 31.80 & 61.2 & 119.2 & 0.244 & 41 &   4 & 11.28 & 5.86 & 38.8 &  5.98 & 20.91 & 11.23 & -21.28 & 2.18 \nl
\nl	
U00908     & N0489      & 35.09 & 20.5 &  43.9 & 0.747 & 80 & 120 & 13.73 & 1.68 &  7.3 &  1.25 & 18.51 & 13.72 & -19.01 & 2.30 \nl
U00914     & N0493      & 32.46 & 58.1 & 113.3 & 0.805 & 90 &  61 & 12.82 & 5.23 & 29.4 &  4.62 & 20.10 & 12.79 & -19.77 & 1.27 \nl
U00947     & N0514      & 34.84 & 53.4 & 114.4 & 0.271 & 44 & 102 & 12.64 & 4.01 & 29.4 &  4.96 & 21.52 & 12.54 & -20.17 & 1.71 \nl
U00952     & N0518      & 37.73 & 26.3 &  75.0 & 0.650 & 72 &  97 & 14.37 & 2.37 & 16.5 &  3.01 & 21.29 & 14.32 & -18.56 & 2.48 \nl
U00966\tablenotemark{f}  & N0520
                        &  \no  &  \no & \no   & \no   &\no & \no &  \no  &  \no &  \no &  \no  & \no   &  \no  &   \no  & \no  \nl
\nl	
U00970     & N0522      & 37.97 & 40.1 &  86.0 & 0.835 & 90 &  33 & 14.14 & 3.35 & 19.4 &  3.57 & 20.32 & 14.09 & -18.81 & 2.51 \nl
U00982     & N0532      & 33.41 & 57.5 & 112.0 & 0.729 & 79 &  29 & 13.66 & 3.65 & 24.1 &  3.91 & 21.04 & 13.60 & -19.02 & 2.44 \nl
U01014     & \no        & 29.94 & 22.8 &  44.3 & 0.091 & 25 & 174 & 14.83 & 1.33 &  9.8 &  1.42 & 21.53 & 14.79 & -17.59 & 1.00 \nl
U01102     & \no        & 27.42 & 28.0 &  37.2 & 0.184 & 36 &   0 & 14.43 & 1.48 &  8.7 &  1.15 & 20.41 & 14.37 & -17.82 & 0.67 \nl
U01104     & \no        & 11.15 & 13.3 &  50.5 & 0.399 & 54 &   6 & 14.49 & 1.45 &  9.4 &  0.51 & 20.96 & 14.49 & -15.75 & 2.31 \nl
\nl	
U01133     & \no        & 27.55 & 32.6 &  63.5 & 0.301 & 46 & 177 & 15.52 & 1.96 & 57.2 &  7.63 & 24.88 & 14.43 & -17.77 & \no  \nl
U01149\tablenotemark{g} & N0628 
			& 10.61 & 81.3 & 158.5 & 0.217 & 39 & 148 & 10.47 &10.26 & 77.6 &  3.99 & 21.52 & 10.38 & -19.75 & 1.55 \nl
U01192     & N0658      & 41.49 & 44.0 & 103.8 & 0.503 & 62 &  28 & 13.27 & 3.42 & 31.0 &  6.24 & 22.41 & 13.17 & -19.92 & 1.71 \nl
U01195     & \no        & 12.03 & 64.4 & 125.5 & 0.799 & 88 &  48 & 13.70 & 4.57 & 31.9 &  1.86 & 21.29 & 13.62 & -16.78 & 1.36 \nl
U01200     & \no        & 12.45 & 17.5 &  66.5 & 0.451 & 58 & 168 & 14.13 & 1.90 & 11.8 &  0.71 & 20.77 & 14.13 & -16.35 & 1.48 \nl
\nl	
U01201     & N0660      & 13.02 & 79.7 & 128.4 & 0.588 & 68 &  23 & 12.16 & 8.00 & 83.5 &  5.27 & 22.83 & 11.96 & -18.61 & 2.21 \nl
U01270\tablenotemark{e} & N0676
                        &  \no  &  \no & \no   & \no   &\no & \no &  \no  &  \no &  \no &  \no  & \no   &  \no  &   \no  & \no  \nl
U01304     & N0693      & 22.11 & 43.2 &  92.7 & 0.597 & 69 & 104 & 13.33 & 3.15 & 27.1 &  2.91 & 22.17 & 13.26 & -18.46 & 2.24 \nl
U01356     & N0718      & 24.23 & 35.3 &  91.6 & 0.137 & 30 &  23 & 12.63 & 3.04 & 22.2 &  2.60 & 21.53 & 12.57 & -19.35 & \no  \nl
U01463     & N0770      & 34.49 & 22.4 &  39.6 & 0.294 & 46 &  10 & 14.45 & 1.20 & 10.8 &  1.80 & 22.34 & 14.42 & -18.27 & \no  \nl
U01466\tablenotemark{h} & N0772
		        & 34.71 & 75.1 & 133.1 & 0.421 & 56 & 124 & 11.25 & 8.73 & 72.4 & 12.18 & 22.01 & 11.14 & -21.56 & 1.76 \nl
\nl
II. \ Fainter & Galaxies: \nl
Z409-018   & \no        & 19.93 & 12.6 &  26.9 & 0.479 & 60 &   6 & 15.40 & 0.98 &  6.2 &  0.60 & 20.89 & 15.36 & -16.14 & 1.51 \nl
Z409-040   & \no        &  9.47 & 11.5 &  18.5 & 0.400 & 54 & 135 & 15.67 & 0.71 &  4.8 &  0.22 & 21.19 & 15.63 & -14.25 & 2.33 \nl
U00494     & \no        & 27.93 & 15.2 &  29.7 & 0.585 & 68 &  95 & 15.21 & 1.03 &  5.1 &  0.69 & 19.42 & 15.19 & -17.04 & 1.40 \nl
U00634     & \no        & 31.38 & 28.4 &  55.2 & 0.567 & 66 &  34 & 15.25 & 2.08 & 20.1 &  3.06 & 22.63 & 15.03 & -17.45 & 1.15 \nl
U00871     & \no        & 30.36 & 14.6 &  28.4 & 0.271 & 44 & 133 & 16.40 & 1.00 & 11.7 &  1.72 & 23.20 & 16.10 & -16.31 & 1.61 \nl
\nl
U00882     & \no        & 32.67 & 24.9 &  44.1 & 0.404 & 55 &  82 & 15.72 & 1.42 & 12.3 &  1.96 & 22.23 & 15.56 & -17.01 & 1.50 \nl
Z411-038   & \no        & 36.84 & 12.4 &  24.1 & 0.229 & 40 &  11 & 15.89 & 0.85 &  7.7 &  1.37 & 22.42 & 15.78 & -17.05 & 1.54 \nl
F0120+0835\tablenotemark{i} & \no    
			& 31.26 & 18.0 &  29.0 & 0.015 & 10 & 149 & 15.83 & 0.88 &  6.1 &  0.92 & 21.30 & 15.78 & -16.70 & 1.82 \nl
Z411-042   & \no        & 37.62 & 24.9 &  40.0 & 0.466 & 59 & 153 & 15.33 & 1.21 & 11.7 &  2.13 & 22.58 & 15.27 & -17.61 & 0.74 \nl
U00964     & \no        & 38.09 & 19.7 &  31.8 & 0.864 & 80 &  90 & 15.40 & 1.12 &  4.7 &  0.86 & 18.17 & 15.38 & -17.52 & 1.06 \nl
F0128+0424 & \no        & 28.68 & 10.4 &  24.6 & 0.336 & 49 & 133 & 16.27 & 0.75 &  5.2 &  0.72 & 21.29 & 16.24 & -16.05 & 1.46 \nl
\tablenotetext{a}{Photometry may be affected by bright stars nearby.}
\tablenotetext{b}{Local Group dwarf galaxies.  Not observed.}
\tablenotetext{c}{The exposure time is only 60 sec due to a bright foreground star.}
\tablenotetext{d}{Extremely edge-on galaxies. No surface photometry.}
\tablenotetext{e}{No data.}
\tablenotetext{f}{An interacting galaxy pair.  No surface photometry.}
\tablenotetext{g}{The galaxy occupies most of the CCD field and the sky subtraction may be less accurate.}
\tablenotetext{h}{The disk fit is performed over a range of radii where spiral arms of the galaxy are prominent.}
\tablenotetext{i}{= MCG+01-04-042.}
\enddata
\end{deluxetable}
\newpage

\begin{deluxetable}{ccccc}
\tablenum{2}
\tablewidth{0pt}
\tablecaption{Sample Selection Sensitivity Indicators for Face-on Disks}
\tablehead{
\colhead{$\mu_0^c$}  & \colhead{$r_s$(min.)} & 
\colhead{$\Gamma(-17.5)/\Gamma^s_{max}$} & 
\colhead{$h_s(-17.5)$}  & \colhead{$h_s(\Gamma^s_{max})$} \\
\colhead{(mag$\,$arcsec$^{-2}$)}  & \colhead{(arcsec)} & \colhead{} &
\colhead{(kpc)}  & \colhead{(kpc)}}
\startdata
25  &  90   &   0.33   &   6.1   &  18.9 \nl
24  &  32   &   0.58   &   3.9	 &   6.7 \nl
23  &  20   &   0.58   &   2.4	 &   4.2 \nl
22  &  13   &   0.57   &   1.5	 &   2.7 \nl
21  &  8    &   0.60   &   1.0	 &   1.6 \nl
20  &  6    &   0.50   &   0.6	 &   1.3 \nl
19  &  5    &   0.37   &   0.4	 &   1.1 \nl
18  & 4.5   &   0.26   &   0.2	 &   1.0 
\enddata
\end{deluxetable}

\begin{references}

\reference{} Allen, R. J., \& Shu, F. H. 1979, \apj, 227, 67
\reference{} Bothun, G. D., Schombert, J. M., Impey, C. D., \& 
             Schneider, S. T. 1990, \apj, 360, 427
\reference{} Bothun, G., Impey, C., and McGaugh, S. 1997, \pasp, 109, 745
\reference{} Briggs, F. H. 1997, \apj, 484, L29
\reference{} Burstein, D., Haynes, M. P., \& Faber, S. M. 1991, Nature,
	     353, 515
\reference{} Byun, Y.-I. 1993, \pasp, 105, 993
\reference{} Cornell, M. E., Aaronson, M., Bothun, G., \& Mould, J. 1987, 
	     \apjs, 64, 507
\reference{} Dalcanton, J. J., Spergel, D. N., \& Summers, F. J. 1997a, \apj, 482, 659
\reference{} Dalcanton, J. J., Spergel, D. N., Gunn, J. E., Schmidt, M., \& 
             Schneider, D. P. 1997b, \aj, 114, 635
\reference{} Davies, J. I. 1990 \mnras, 244, 8
\reference{} de Jong, R. S. 1996, \aap, 313, 45
\reference{} de Vaucouleurs, G., de Vaucouleurs, A., Corwin, H. G., Buta, 
	     R. J., Paturel, G., \& Foueu\'e, P. 1991, Third Reference 
	     Catalogue of Bright Galaxies (New York: Springer) (RC3)
\reference{} Disney, M. J. 1976, \nat, 263, 573
\reference{} Disney, M. J., \& Phillipps, S. 1983, \mnras, 205, 1253
\reference{} Disney, M. J., Davies, J., \& Phillipps, S. 1989, \mnras, 239, 939
\reference{} Freeman, K. C. 1970, \apj, 160, 811
\reference{} Giovanelli, R., Haynes, M. P., Salzer, J. J., Wegner, G., Costa, L. N.,
             Freudling, W. 1994, \aj, 107, 2036
\reference{} Gunn, J. E., \etal 1987, Opt.~Engineering, 26, 779
\reference{} Hoffman, G. L., Dickey, J., Lu, N. Y., \& Fromhold-Treu, R. 1996, 
 	     \apj, 473, 822
\reference{} Johnson, H. L., \& Morgan, W. W. 1953, \apj, 117, 113
\reference{} Kent, S. M. 1985, \apjs, 59, 115
\reference{} Kormendy, J. 1977, \apj, 217, 406
\reference{} Landolt, A. U. 1983, \aj, 88, 439
\reference{} Lu, N. Y., Hoffman, G. L., Groff, T., Roos, T., \& Lamphier, C.
             1993, \apjs, 88, 383 (Paper~I)
\reference{} Lu, N. Y., Salpeter, E. E., \& Hoffman, G. L. 1994, \apj, 426, 473
\reference{} Malmquist, K. G. 1920, Medd.~Lunds Astron.~Obs., Ser.~II, No.~22
\reference{} McGaugh, S. S. 1996, \mnras, 280, 337
\reference{} McGaugh, S. S., Bothun, G. D., \& Schombert, J. M. 1995, \aj, 110, 573
\reference{} Nilson, P. 1973, Uppsala General Catalogue of Galaxies, Uppsala
	     Astron. Obs. Ann., 6 (UGC)
\reference{} Pierce, M. J., \& Tully, R. B. 1988, \apj, 330, 579
\reference{} Schombert, J. M., Bothun, G. D., Schneider, S. E., McGaugh, S. S. 1992,
             \aj, 103, 1107
\reference{} Sprayberry, D., Impey, C. D., \& Irwin, M. J. 1996, \apj, 463, 535
\reference{} Szomoru, A., Guhathakurta, P., van Gorkom, J. H., 
	     Knapen, J. H., Weinberg, D. H., \& Fruchter, A. S. 1994, 
	     \aj, 108, 491
\reference{} Tully, R. B., \& Fouqu\'e, P. 1985, \apjs, 58, 67
\reference{} Tully, R. B., \& Verheijen, M. A. W. 1997, \apj, 484, 145
\reference{} Tully, R. B., \& Verheijen, M. A. W., Pierce, M. J., Huang, J.-S.,
	     Wainscoat, R. J. 1996, \aj, 112, 2471
\reference{} Valentijn, E. A. 1990, Nature, 346, 153
\reference{} Valentijn, E. A. 1994, MNRAS, 266, 614
\reference{} van der Kruit, P. C. 1987, \aap, 173, 59
\reference{} Xu, C., \& Buat, V. 1995, \aap, 293, L65
\end{references}
\end{document}